\documentclass[a4paper,11pt]{article}
\pdfoutput=1
\usepackage{jheppub}
\usepackage{amsmath,amsthm,amssymb,braket,color,verbatim,adjustbox,multirow,enumerate,textcomp,gensymb,subfigure,graphicx,diagbox,pifont,tabulary,booktabs,stackengine,epstopdf,dcolumn,makecell,listings,bm,mathtools,autobreak,slashed,float,array,colordvi,xfrac}
\usepackage{hyperref}
\hypersetup{colorlinks,bookmarksopen,bookmarksnumbered,citecolor=[rgb]{0,0.35,0.5},linkcolor=[rgb]{0,0.35,0.5},urlcolor=[rgb]{0,0.35,0.5},pdfstartview=FitH,linktocpage}

\definecolor{eprintLinks}{rgb}{0,0.35,0.5}
\definecolor{journalLinks}{rgb}{0,0.35,0.5}
\newcommand{\MYhref}[3][blueLinks]{\href{#2}{\color{#1}{#3}}}

\def\Mp{M_{P}}
\def\g{\gamma}
\def\O{{\cal O}}
\def\pv{{\bf p}}
\def\pn{\lvert\pv\rvert}
\def\kv{{\bf k}}
\def\kn{\lvert\kv\rvert}
\def\M{{\cal M}}
\def\Msq{\lvert\M\rvert^{2}}
\def\Msqa{\overline{\Msq}}
\def\kp{k_{p}}
\def\sd{\textrm{d}}
\def\Ms{M_{s}}
\def\gs{g_{s}}
\def\Md{M_{D}}
\def\sdla{\langle\!\langle}
\def\sdra{\rangle\!\rangle_{D}}
\def\wa{\overline{\omega}}
\def\pa{\overline{p}}
\def\Ea{\overline{E}}

\def\ka{\overline{k}}
\def\nd{n_{D}}
\def\ls{\ell_{s}}
\def\alp{\alpha^{\prime}}
\def\uv{{\bf u}}
\def\un{\lvert\uv\rvert}
\def\D{\mathfrak{d}}
\def\Dn{\mathfrak{d}_{\nu}}
\def\Dg{\mathfrak{d}_{\gamma}}
\def\E{{\cal E}}
\def\i{\textrm{i}}
\def\e{\textrm{e}}
\def\Bdla{\Bigl\langle\!\!\Bigl\langle}
\def\Bdra{\Bigr\rangle\!\!\Bigr\rangle_{D}}
\def\N{{\cal N}}
\def\I{{\cal I}}

\renewcommand{\thefootnote}{\fnsymbol{footnote}}

\title{\boldmath Constraints to Lorentz violation and ultrahigh-energy electrons in D-foamy space-times\footnote{Published as JHEP 10 (2025) 216, \url{https://doi.org/10.1007/JHEP10(2025)216}}}

\author[a]{Chengyi Li}\emailAdd{lichengyi@pku.edu.cn}
\affiliation[a]{School of Physics, Peking University, Beijing 100871, China}
\author[b,a]{and Bo-Qiang Ma\footnote{Corresponding author.}}\emailAdd{mabq@pku.edu.cn}
\affiliation[b]{School of Physics, Zhengzhou University, Zhengzhou 450001, China}

\abstract{We entertain the constraints that the absence of vacuum Cherenkov radiation of ultrahigh-energy electrons inferred from LHAASO observations of the Crab Nebula can impose on generic models in which Lorentz symmetry of the particle vacuum is violated, as established by some recent studies in \href{https://doi.org/10.1016/j.physletb.2022.137034}{\emph{Phys. Lett. B} {\bf 829} (2022) 137034}; \href{https://doi.org/10.1016/j.physletb.2022.137536}{{\bf 835} (2022) 137536}; \href{https://doi.org/10.1103/PhysRevD.108.063006}{\emph{Phys. Rev. D} {\bf108} (2023) 063006}. We demonstrate in the present paper, that implementing a phenomenological approach to the Lorentz violation, the rates of this vacuum process are substantial such that one is justified in deriving bounds on the violation scales from simple threshold analysis just as these works did. Albeit such results are likely effective then, they do not apply in the same form among scenarios. Specifically, we show that these Cherenkov constraints are naturally evaded in models of space-time foam inspired from~(supercritical) string theory, involving D-branes as space-time defects in a brane-world scenario, in which subluminous energy-dependent refractive indices of light have been suggested. We examine here two specific foam situations and find for both cases~(though, for different reasons) the potentiality that charged quanta such as electrons do \emph{not} radiate as they pass through the gravitational vacuum `medium' despite moving faster than photons.}


\arxivnumber{2505.06121}

\begin{document}
\maketitle
\flushbottom

\renewcommand{\thefootnote}{\arabic{footnote}}


\section{\label{sec:intro}Introduction and summary}

The discovery of PeV photons reported by the Large High Altitude Air Shower Observatory~(LHAASO) from PeVatrons~\cite{LHAASO:2021gok,Li:2021tcw,LHAASO:2023uhj} including the ancient Crab Nebula~\cite{LHAASO:2021cbz}, kicks off a new era in astronomy, providing a unique opportunity to probe the potential breakdown of Lorentz invariance~(LI)~\cite{Li:2025stf,AlvesBatista:2023wqm,He:2022gyk} arising from quantum gravity~(QG) analysis. High-energy violation of this symmetry~(i.e., Lorentz violation~(LV)) leads to particle-energy~($E$) dependent physics, that is suppressed by the gravitational scale $M\sim\Mp\approx 10^{19}$~GeV where $\Mp$ is the Planck mass. As a result, for extreme energies reached in astrophysical processes, its consequences could be identified in the current observations. The question arises, then, what information we can get from LHAASO experiments of ultrahigh-energy~(UHE) $\g$-rays and particularly, from $\sim 1.1$~PeV Crab photon. Recent studies~\cite{He:2022jdl}, carrying further a previous analysis~\cite{Li:2022ugz} by the present authors, pointed out that the~(indirect) observations of UHE electrons assumed to produce the PeV emission of the Crab severely limit \emph{minimally} suppressed LV \emph{dispersion} corrections, i.e., of $\O(E/M)$, to both electrons and photons.

Such bounds were placed via considerations of the vacuum Cherenkov process, $e\rightarrow e\g$, which is forbidden in LI quantum electrodynamics~(QED) since the speed of a fast moving charge is always less than that of photon propagation. With LV it can be~\cite{Coleman:1997xq} an important energy-loss channel for the most high-energetic particle. As often adopted in the literature to parameterize LV~(see~\cite{Li:2025stf}), also in the same spirit as~\cite{He:2022jdl,Li:2022ugz}, \emph{a priori} most phenomenological studies converge in assuming a generic, nonstandard dispersion relation:
\begin{equation}\label{eq:gdr}
E_{I}^{2}-\pv_{I}^{2}\bigl(1+\delta^{(I)}\bigr)+\ell_{I,n}\lvert\pv_{I}\rvert^{n+2}=m_{I}^{2},\quad n\geq 1,
\end{equation}
where $(E_{I},\pv_{I})$ is the four-momentum associated with a given particle $I$ of~(rest) mass $m_{I}$, and the modulus of the spatial momentum is sometimes denoted, below, by $p\coloneqq\pn$. [N.B.: the notation follows prior studies of~\cite{He:2022jdl,Li:2022ugz}.] $\ell_{I,n}\coloneqq s_{I,n}/(E_{\textrm{LV}I,n})^{n}\equiv -\xi_{I,n}/M^{n}$ defines the scale of the $n$th-order LV physics beneath,\footnote{Here, the dimensionless LV parameters $\xi_{I,n}$ are defined to simplify some of the expressions below, albeit they may be absorbed into redefinition of the  mass $M$, and, then give the~(positive) LV energy scale, $E_{\textrm{LV}I,n}$ by keeping only the LV sign factor $s_{I,n}(=+1$ or $-1)$ explicitly.} while $\delta^{(I\,=\,e)}$ parameterizes the strength of $\O(p^{2})$ electron LV that has been constrained to be extremely small, $\lvert\delta\rvert<10^{-20}$~\cite{Li:2022ugz}, implying that for astrophysically relevant processes, $\O(1/M^{n})$ terms dominate.

For example, in a linear LV model~($n=1$, in which case this suffix $n$ is omitted) often considered, modified dispersion~(\ref{eq:gdr}) induces an energy-dependent velocity of propagation {\it in vacuo} for both the electromagnetic waves, $\Delta v_{\g}\coloneqq\partial E_{\g}/\partial p_{\g}-1\simeq -\,(\ell E)_{\g}$ and electron waves, $\Delta v_{e}\simeq -[m_{e}^{2}/(2E_{e}^{2})+(\ell E)_{e}]$, assuming vanishing $\delta$, with $s_{\g/e}=+1$~($-1$) for subluminal~(superluminal) propagation relative to low-energy photons that are not supposed to be affected significantly by the LV effects, and that will travel at the conventional speed of light $c$~(already set to be unity).\footnote{\label{fn:1}Such nontrivial propagation properties can be found from concrete models. For example within effective field theory approach to modified QED, models with operators of dimension-5 are compatible with the case of exhibiting birefringence~(i.e., having~\cite{Myers:2003fd} helicity-dependent $\ell_{\g/e}$) while~(Liouville~\cite{Ellis:1992eh}) string inspired \emph{foamy} space-time `\emph{medium}'~\cite{Mavromatos:2002re} based on D-brane recoil~\cite{Ellis:1999uh} dictates a subluminal refraction for photons~(and \emph{not} electrons, c.f. $\ell_{e}=0$ at tree-level), see section~\ref{sec:sstf}.} Aside from such propagation effects \emph{in vacuo} there is the possibility of peculiar novel~(threshold) reactions beyond the standard model, like $e\rightarrow e\g$, provided the energy–momentum conservation relation is kept intact, an assumption usually adopted in a purely phenomenological formulation of LV, e.g.,~(\ref{eq:gdr}), owing to a lack of any particular prior knowledge.

Depending on energies of the emitted photons the process can be either `soft' or `hard' and the so-called inverse-Compton~(IC)--Cherenkov bound, a sort of LV constraint we shall be concerned with, is then provided by electrons and positrons of some energy $E_{e}$ inferred via observations of $\g$-rays of energy $\leq E_{e}$ from the Crab Nebula whose radiation $\gtrsim 1$~GeV is explained by IC scattering~\cite{Atoyan:1996kqd}, a process hardly affected by LV~(see appendix of~\cite{Li:2022ugz}). As the rate of $e\rightarrow e\g$ is orders of magnitude greater than that of IC scattering, this reaction must not take place for these charges~\cite{Jacobson:2005bg,Li:2022ugz}. In case there are Lorentz violations present in \emph{both} fermions and photons, i.e., having nonzero $\ell_{e}$ and $\ell_{\g}$, the constraint depends on both parameters and will form an excluded region in parameter space. Especially in~\cite{He:2022jdl} this was done thoroughly, as mentioned, using recent first observation of PeV Crab radiation which requires at least 2.3~PeV long living electrons~\cite{LHAASO:2021cbz,Li:2022ugz}.

For our purposes in this paper, we do not discuss the bounds in this very general case, but give instead a brief summary of the results that are the ones obtained as special cases, i.e., the cases with electrons and photons subject to LVs, \emph{respectively}~(for details, see~\cite{He:2022jdl}, if interested). Specializing to these two cases will be sufficient to capture kind of core feature of this reaction, and is also reasonable, at least as long as one notices the concrete examples provided by brane theories, where only certain~(neutral) particles may `feel' the \emph{space-time foam} effect, as already mentioned in fn.~\ref{fn:1} and will be discussed below. For a more detailed discussion of vacuum Cherenkov kinematics, in general terms, as well as its dependence on the hierarchies among the LV parameters for the electron~($\xi_{e}\eqqcolon\eta$) and of photon~($\xi_{\g}\eqqcolon\xi$) see, e.g., refs.~\cite{Jacobson:2005bg,Jacobson:2002hd,Konopka:2002tt} and~\cite{Mattingly:2005re}.
\begin{itemize}
\item If the photon dispersion is unaffected~($\xi=0$), then the reaction occurs with emission which is very soft once the electron group velocity approaches the low energy velocity of light, $\Delta v_{e}(E_{e,th})\approx 0$ which gives,
\begin{equation}\label{eq:sct}
E_{e,th}\simeq\Bigl(\frac{m_{e}^{2}M}{2\eta}\Bigr)^{1/3}.
\end{equation}
It sets what is called \emph{soft} Cherenkov threshold for emitting a zero energy photon, the most studied case in the literature~(e.g.,~\cite{Li:2022ugz}) and is only valid for $\eta>0$, i.e., fermions subject to the \emph{superluminal} LV correction. Provided the Crab Nebula operates as an electron PeVatron~\cite{LHAASO:2021cbz}, $E_{e,th}\geq 2.3$~PeV therefore, $\eta\lesssim 1.3\times 10^{-7}$, or equivalently~\cite{Li:2022ugz}: for $s_{e}=-1$, $E_{\textrm{LV}e}>(7.7\times 10^{6})\times\Mp$ which is very stringent.
\item There is also the configuration of $(\eta$, $\xi)$ which involves emission of a hard photon~\cite{Jacobson:2002hd,Konopka:2002tt} that can carry off half the initial energy in the case of keeping fermion dispersion intact~($\eta=0$) and having \emph{subluminal} LV in photon dispersion. In this case this \emph{hard} threshold can be computed, using the `threshold theorem'~\cite{Mattingly:2002ba}, as
\begin{equation}\label{eq:hct}
E_{e,th}\simeq\Bigl(-\frac{4m_{e}^{2}M}{\xi}\Bigr)^{1/3},\quad\xi<0.
\end{equation}
While it can also be derived with the fact that at threshold, the group velocity of the outgoing electron is equal to that of photons~(i.e., $\Delta v_{e}(E_{e,th}/2)\approx\Delta v_{\g}(E_{e,th}/2)$), this is not the case in more general case with $\eta\neq 0$; see~\cite{Jacobson:2005bg}. Electrons with $E_{e}>2$~PeV from LHAASO~\cite{LHAASO:2021cbz} imply $\xi>-10^{-6}$~\cite{He:2022jdl}.\footnote{Although the cross-channel of $e\rightarrow e\g$~(c.f. `\emph{photon decay}' $\g\rightarrow ee^{+}$) was also used in~\cite{He:2022jdl} to set bounds it can only constrain subluminal modification of electron dispersion and superluminal photons, both of which are absent in brane model(s) we shall work with. It is for this reason we ignore at this stage $\g$-decay~(which does not occur in such models) except for some discussion in concluding~(section~\ref{sec:concl}).}
\end{itemize}

Quite interestingly, the absence of hard Cherenkov threshold severely constrains \emph{subluminal} photon dispersions with \emph{linear} LV modification, an effect compatible with a previous hint of \emph{light-speed variation} in gamma-ray burst~(GRB) photon flight time delays~\cite{Shao:2009bv}, that require $E_{\textrm{LV}\g}\sim 10^{17}$~GeV, leading to a saturation of the latter. This motivates us to reconsider the bound from~(\ref{eq:hct}). In fact, only if we confirm that electrons lose energy extremely quickly through this process can we reliably apply the LV bounds derived from the threshold considerations above. To date, this issue has been examined only for specific models of LV~\cite{Altschul:2007tn,Klinkhamer:2008ky,Anselmi:2011ae,Rubtsov:2012kb}, while the vacuum Cherenkov rate within purely phenomenological approaches, such as those described by~(\ref{eq:gdr}), remains unclear. In the first part of this paper, we aim to briefly discuss this topic. Above the threshold, as we will demonstrate, the reaction rate is significant, allowing straightforward kinematic threshold analyses, as seen in refs.~\cite{He:2022jdl,Li:2022ugz}, to yield reliable constraints.

However, at the kinematic level, permitting electron Cherenkov effect \emph{in vacuo} requires two conditions which may need re-evaluation in certain QG frameworks: \textbf{i)} the adoption of LV-modified dispersion relations in eq.~(\ref{eq:gdr}), and \textbf{ii)} the preservation of energy--momentum conservation. If either condition fails, `\emph{electron decay}' through such an anomalous channel could be ruled out~(much like in the standard model, though for entirely different reasons), despite the presence of LV in the electron and/or photon sectors. This scenario may actually describe certain~(solitonic) string/brane models of space-time foam, which we will examine more closely later in this work.

To this end, we shall first analyze the phenomenological leading-~(linear-)order modification of the rate of vacuum Cherenkov radiation due to the presence of general dispersion relations~(\ref{eq:gdr}) allowed by hard Lorentz violations. By a direct calculation, we confirm that the rapidity of the process in both soft and hard emission cases provides justification for a purely kinematical~(threshold) analysis; IC--Cherenkov bounds~\cite{He:2022jdl,Li:2022ugz} just revisited are thus effective, as are those set~\cite{He:2022jdl} to the subluminal linear term in photon's LV dispersions~(\ref{eq:gdr}) in this generic setup. However these strong constraints \emph{are not always} applicable regarding particular QG proposal. Specifically, considering the fact that charged particles cannot be refracted by the Lorentz-violating vacuum associated with D-brane foam models of space-time~(`\emph{D-foam}')~\cite{Ellis:1999uh,Ellis:2000sf,Ellis:2004ay,Mavromatos:2005bu,Ellis:2008gg,Ellis:2009vq,Li:2023wlo,Li:2021gah,Li:2024crc}, we suggest two mechanisms, through which the UHE electrons emit \emph{no} Cherenkov radiation, despite traveling faster than photons of comparable energies in such backgrounds where photons are slowed down due to the medium effects, with which the aforementioned light-speed variation has been explained~\cite{Li:2021gah}.

\section{\label{sec:crc}Cherenkov emission and rate: an illustrative case study}

In this section we study the rates of vacuum Cherenkov radiation with a phenomenological description of LV via deformed particle dispersions of the form~(\ref{eq:gdr}). Although, one might argue that without adopting a full-fledged LV model not much can really be predicted with confidence, nevertheless for our purposes in this section, this will not be quite relevant and, it suffices to demonstrate the results according to the formalism of eq.~(\ref{eq:gdr}) we considered. We shall use conservation law as well, as commonly used in the phenomenological analysis, as in~\cite{He:2022jdl,Li:2022ugz}, while by itself this needs support from specific theories. Clearly in this context possible contributions of LV to interaction dynamics~(if any) cannot be taken into account, as the analysis is just based on dispersions. A bit of clarification of this is presented in the following subsection,~\ref{sec:ncr}. In subsection~\ref{sec:dar}, by a direct calculation, we briefly study photon emission process of energetic electrons, presenting illustrative formulae for the decay width and energy loss rate valid for both soft and hard radiated photons. Then, it is not difficult to observe that the process $e\rightarrow e\g$ is extremely efficient, hence we will be justified in using threshold values to constrain Lorentz violation. In section~\ref{sec:com}, we compare our result with some analytical ones derived in concrete models of QED~(particularly the ones mentioned in the introduction), reproducing some of their results.

\subsection{\label{sec:ncr}A note on the calculation of rates in Lorentz violation contexts}

We consider it as instructive to remind the reader that employing the microscopic language of rates, LV effect in a complete treatment is threefold. First, of particular relevance for us, deviations of phase space of particle kinematics. Second, modifications of wave functionals of external legs and the third, changes in the vertices and propagators. Scattering matrices may then get modified for the latter two types of corrections~(compared to the standard LI case). However, considering the modified form of the matrix element $\M$ would require the precise knowledge of the LV dynamics, for example~(in the case of field-theoretical models) the novel Feynman~(and, spin sum) rules, due to extra terms present in the corresponding Lagrangian, which is obviously absent in our phenomenological setup~(\ref{eq:gdr}). Of course, one can plausibly assume at least in the low energy limit, the amplitude squared~(summed and averaged over spins and/or polarizations):
\begin{equation}\label{eq:msa}
\overline{\sum}\lvert\M^{\textrm{LV}}(\{p\})\rvert^{2}=\overline{\lvert\M(\{p\})\rvert^{2}}\,\bigl[1+\O(E/M)\bigr],
\end{equation}
where, $\Msqa\equiv\overline{\sum}\Msq$ is the standard matrix element of a process in the LI QED and $\{p\}$ denotes its dependence on the particle momenta. While it has been shown that at least for certain models of LV QED~(e.g.,~\cite{Rubtsov:2012kb}) all these effects contribute equally in magnitude and should be invoked simultaneously, in the following, for reasons just stated, we focus on the first element, i.e., phase space correction, the only effect directly related to~(\ref{eq:gdr}), standing out to us as a `semi-microscopic' treatment of the problem.

Additionally, this section focuses solely on illustrating the calculation of the Cherenkov rate to support the estimation of bounds derived from kinematic thresholds. This complements recent studies in~\cite{He:2022jdl,Li:2022ugz}, and for consistency, we adopt the same theoretical framework as those studies. Consequently, we neglect potential effects arising from~(\ref{eq:msa}), assuming LI matrix element. Nonetheless as mentioned above, such reaction rates have been calculated in many earlier works, to which we refer the reader for details, and where one considers in general a concrete theory so that the dynamical effects~(c.f.~(\ref{eq:msa})) can be treated properly. For example, modifications of phase-space integrals and to the functional form of external states were studied in~\cite{Jacobson:2005bg} in the context of QED with dimension-5 field operators of which the dispersion induced is of $n=1$ type~\cite{Myers:2003fd} thus bears a similarity with the one considered here. Later, spin sums over external states for models with operators of higher dimensions were noted in~\cite{Anselmi:2011ae}, which built on earlier calculations of~\cite{Klinkhamer:2008ky}\footnote{Albeit severely constrained via the absence of detectable birefringence the complement of this model~(in view of CPT status of LV terms), the Chern--Simons one, has been studied extensively in the literature and $e\rightarrow e\g$ \emph{in vacuo} were also explored in this context, both microscopically~\cite{Kaufhold:2007qd} and macroscopically~\cite{Lehnert:2004hq}.} where the model is restricted to modified Maxwell theory. A systematic analysis incorporating dynamical aspects of LV for dimension-4 and -6 terms was carried out soon after in ref.~\cite{Rubtsov:2012kb} and it is the procedure for evaluation of rates described therein that we shall follow in our study. Of course, there also exist attempts similar to us, i.e., to consider radiation rates for the general case which converges quicker and simpler to phenomenology via~(\ref{eq:gdr}) and is not~(necessarily) bounded to a particular model; see for instance~\cite{Martinez-Huerta:2017ulw}, where the treatment is more general and, thus, technically more involved. However therein \emph{a priori} the effect of LV in fermions is ignored; this is not the case here along the next subsection.

\subsection{\label{sec:dar}Decay width for anomalous Cherenkov process of charges and the rate}

We will describe now why one can be justified in utilizing kinematical threshold conditions mentioned previously, for the purposes of Lorentz tests. With the simplifying assumptions above, we work directly with the QED vertex, $e^{\mp}(p)\rightarrow e^{\mp}(p^{\prime})\g(k)$, where, for quantities in the parentheses, the 4-momenta, we leave out suffixes $e$, for clarity, and further use photon wave 4-vector $k^{\mu}$. Writing out the dispersion relations fulfilled by the electron and photon, respectively, as shown in~(\ref{eq:gdr}), yields,
\begin{equation}\label{eq:rdr}
E^{2}=m_{e}^{2}+\pv^{2}\Bigl[1+\frac{1}{2}(\delta+\eta\pn/M)\Bigr]^{2},\quad E_{\g}^{2}=\kv^{2}+\xi k^{3}/M,
\end{equation}
where the residual rotation invariance implicit in eq.~(\ref{eq:gdr}) has been used.\footnote{We remind the reader, that this simplifying choice is used implicitly in most phenomenological analyses of LV, say, in~\cite{He:2022jdl,Li:2022ugz}, as is also the cases of some noneffective field-theoretical QG models including the ones based on brane theories discussed below; otherwise spatial anisotropies enter~(\ref{eq:rdr}) via direction-$i$-dependent $\eta_{i},\xi_{i}$~(for implications of broken O(3) symmetry see, e.g.,~\cite{Mattingly:2005re}).} For the energy well above the threshold of interest here, the lepton masses can be safely neglected so that the electron/positron dispersion further simplifies to $E^{2}\simeq \pv^{2}(1+\delta+\eta p/M)$, and for the outgoing particle $E^{\prime 2}\simeq \pv^{\prime 2}(1+\delta+\eta p^{\prime}/M)$. We emphasize that there is no helicity or spin dependence of the LV terms in our oversimplified formalism~(\ref{eq:rdr}), i.e., $\eta$ and $\xi$ are assumed to be \emph{either} positive or negative~(or equivalently, $s_{e/\g}$ takes only one value between $-1$ and $+1$). Otherwise, for example, there would be photon birefringence which has been severely constrained by cosmological measurements~\cite{Li:2025stf,AlvesBatista:2023wqm,He:2022gyk}.

We start with some kinematical considerations, basing our analysis in a frame in which the motion of the initial lepton is along, say, $x^{1}$ coordinate for concreteness and brevity. It allows one to parameterize the 4-momenta of the final states as
\begin{equation}\label{eq:fmp}
p^{\prime\mu}=\bigl(E^{\prime},p(1-\kp),\lvert\pv_{\perp}^{\prime}\rvert,0\bigr),\quad k^{\mu}=\bigl(E_{\g},p\kp,-\lvert\pv_{\perp}^{\prime}\rvert,0\bigr),
\end{equation}
along with $E_{\g}=E-E^{\prime}$. Clearly, up to rotations, it is the most general parameterization satisfying the energy--momentum conservation. Note that the conservation of energy in the zeroth order in the LV parameters, requires the fraction of momentum carried away by the radiated photon to lie in the range:
\begin{equation}\label{eq:emf}
\kp\coloneqq\frac{\kn}{\pn}\sim1-\frac{E^{\prime}}{E}\in [0,\kp^{\max}]\quad (\kp^{\max}\leq 1).
\end{equation}
In a highly relativistic regime that is of interest in astrophysics~(c.f. astroparticle tests~\cite{He:2022jdl,Li:2022ugz} of LV) and, which we make use here, one has $\lvert\pv_{\perp}^{\prime}\rvert\eqqcolon p_{\perp}^{\prime}\ll\pn$. This allows to expand the energies of the particles as follows,
\begin{align}\label{eq:pee}
E&=p+\frac{1}{2}\delta p+\frac{\eta}{2M}p^{2},\quad\ \, E_{\g}\simeq p\kp+\frac{\xi}{2M}p^{2}\kp^{2}+\frac{p_{\perp}^{\prime 2}}{2p\kp},\nonumber\\
E^{\prime}&\simeq p(1-\kp)\Bigl(1+\frac{1}{2}\delta\Bigr)+\frac{\eta}{2M}p^{2}(1-\kp)^{2}+\frac{p_{\perp}^{\prime 2}}{2p(1-\kp)}.
\end{align}

To proceed with some details of the evaluation we next consider the squared amplitude probability for the channel $e\rightarrow e\g$, computed with the standard QED rules. The tree-level matrix element then takes the simple form
\begin{equation}\label{eq:cme}
\M=\bar{u}(p^{\prime})(-e\g^{\mu})u(p)\epsilon^{\ast}_{\mu}(k),
\end{equation}
where $e$ is the elementary charge, associated with the fine structure constant $\alpha_{e}=e^{2}/(4\pi)$, and, $u$ is the Dirac spinor wave-function of leptons, $\epsilon_{\mu}^{\ast}$ the conjugated photon polarization. We note again that although it is not the case for some scenarios within brane-world/string theory discussed later on, modified dispersion relations for the particles may be formulated in local field theory in terms of effective Lagrangian, which in turn will affect the spin sum of particles and their interaction vertices. Clearly as explained, the present analysis cannot deal with such subtleties or model dependencies.

The next step is to sum over the spins and/or polarizations of the final and to average over the initial leptons' helicities in order for obtaining the inclusive rate, as is appropriate for the randomly polarized electrons~(positrons) inside the Crab pulsar wind. In short, the spin-summed and -averaged differential width obeys the standard formula which is valid in the LV contexts~(as its derivation does not assume Lorentz invariance),
\begin{equation}\label{eq:ddw}
\sd\Gamma_{e\rightarrow e\g}=\frac{\Msqa}{2E}\sd\Phi_{2},
\end{equation}
and noting that the radiated 4-momentum per unit time $\sd\dot{p}_{\mu}\sim p_{\mu}\sd\Gamma$ then the corresponding energy loss rate may be defined as
\begin{equation}\label{eq:elrd}
\dot{E}\coloneqq\int (E^{\prime}-E)\sd\Gamma_{e\rightarrow e\g}.
\end{equation}
Here, the squared modulus of the transition amplitude summed and averaged over relevant states, is found from~(\ref{eq:cme}), $\Msqa=4e^{2}(-p\cdot p^{\prime})$, containing information of the dynamics of the decay.\footnote{For the purposes of our discussion below, we used the convention, $\eta_{\mu\nu}=\textrm{diag}[-1,1,1,1]$, the `east coast' metric that is often taken in relativity, gravity and string communities but, less common in particle physics. Books about standard-model field theory based on our choice~\cite{Li:2024crc} of metric do exist; see, e.g.,~\cite{Burgess:2006hbd}, in which with a clear description of how to convert between conventions, the authors imply $[-\bar{\psi}(\slashed{\partial}+m)\psi]$, the Dirac Lagrangian, and the spin sums $\sum_{s}u\bar{u}/v\bar{v}(p,s)=-\i\slashed{p}\pm m$.} The remaining part, the 2-body phase-space measure, describes its kinematics, $\sd\Phi_{2}=(2\pi)^{4}\delta^{(4)}(p^{\prime}+k-p)\sd\Pi_{e}\sd\Pi_{\g}$ with the element $\sd\Pi_{I}$ determined by the conventional relation $2E^{\prime}(p^{\prime})\sd\Pi_{e}=(2\pi)^{-3}\sd^{3}\pv^{\prime}$, $2E_{\g}(k)\sd\Pi_{\g}=(2\pi)^{-3}\sd^{3}\kv$. Using the kinematics defined earlier one can check with ease that the phase space of outgoing particles can be rewritten in terms of the variables $\kp$ and $p_{\perp}^{\prime}$. Therefore one has,
\begin{equation}\label{eq:cdr1}
\Gamma_{e\rightarrow e\g}=\frac{4e^{2}}{16\pi}\frac{1}{2p}J,
\end{equation}
and the phase space integral $J$ surviving from eliminating the $\delta^{(3)}$ function associated with spatial momentum conservation reduces to,
\begin{equation}\label{eq:psi1}
J=\int\frac{\sd\kp\sd p_{\perp}^{\prime 2}}{p\kp(1-\kp)}2\delta\Bigl(F(\kp)-\frac{p_{\perp}^{\prime 2}}{p\kp(1-\kp)}\Bigr)[EE^{\prime}-(\pv\cdot\pv^{\prime})],
\end{equation}
where one introduced the notation:
\begin{equation}\label{eq:lvps}
F(\kp)=\delta\pn\kp-\frac{\xi}{M}\pv^{2}\kp^{2}+\frac{\eta}{M}\pv^{2}(2\kp-\kp^{2}).
\end{equation}
Plugging further the ultrarelativistic kinematics~(\ref{eq:pee}), whose modifications have been kept only up to leading terms in the LV parameters, into the part in the square bracket, coming from the matrix element, the integrand~(\ref{eq:psi1}) becomes,
\begin{align}\label{eq:psi2}
J=2\int\sd p_{\perp}^{\prime 2}\int&\,\sd\kp\frac{1}{\kp(1-\kp)}\delta\Bigl(pF(\kp)-\frac{p_{\perp}^{\prime 2}}{\kp(1-\kp)}\Bigr)\nonumber\\
&\times\Bigl[\delta p^{2}(1-\kp)+\frac{\eta}{2M}p^{3}(1-\kp)(2-\kp)+\frac{p_{\perp}^{\prime 2}}{2(1-\kp)}\Bigr],
\end{align}
which despite its apparently complicated form, is just an elementary integration. Performing the integral over $p_{\perp}^{\prime}$, then, we obtain the differential emission rate:
\begin{equation}\label{eq:cdw}
\frac{\sd\Gamma_{e\rightarrow e\g}}{\sd\kp}=\alpha_{e}\biggl\{\delta\pn\Bigl[1+\frac{\kp}{2}(\kp-2)\Bigr]-\frac{\xi}{2M}\pv^{2}\kp^{3}+\frac{\eta}{2M}\pv^{2}\bigl[1-(\kp-1)^{3}\bigr]\biggr\}.
\end{equation}
One observes that the rate is suppressed by the parameters describing LV. This suppression stems from the phase space integration, which is identically zero in the LI case and remains small as long as LV is small, because the available phase space is very limited for the nearly collinear kinematics realized in the decay.

To obtain the total width we have to work out the final $\kp$-integration. One finds that the domain of integration in $\kp$ is determined by the requirement, that $p_{\perp}^{\prime 2}$ found from the $\delta$ function in eq.~(\ref{eq:psi2}) is positive, i.e.,
\begin{equation}\label{eq:irc}
F(\kp)\geq 0.
\end{equation}
This implies that, while the $\kp$-range is of the standard form~(\ref{eq:emf}), for some $\kp^{\max}$ specified by this condition, it will actually depend on the hierarchies among the LV parameters. For our purpose of showing that the decay is so fast that in particular in the above-mentioned cases of our interest~(i.e., either $\xi$ or $\eta$ is zero) it is practically governed by pure kinematics, classifying all possibilities is not relevant for us. In fact, it suffices to consider the simplest configuration $0\leq\delta$, $0\leq\eta$, $\xi\leq\eta$, which is appropriate for achieving our aim. In this case the integral is over the whole range $0\leq\kp\leq 1$, yielding,
\begin{equation}\label{eq:cdr2}
\Gamma_{e\rightarrow e\g}(\pv)=\alpha_{e}\pn\,\biggl[\frac{2}{3}\delta+\Bigl(\frac{5}{8}\eta-\frac{1}{8}\xi\Bigr)\frac{\pn}{M}\biggr].
\end{equation}
This result is exact with the only approximation being $m_{e}=0$. For one of the most useful characteristic of the process, the rate of the energy decrease by the electron, c.f.~(\ref{eq:elrd}), one multiplies the integrand of $J$ by $-E\kp$ and obtains,
\begin{equation}\label{eq:elr}
\frac{\sd E}{\sd t}\approx -\int\sd\kp\, p\kp\frac{\sd\Gamma_{e\rightarrow e\g}}{\sd\kp}=-\alpha_{e}\pv^{2}\biggl[\frac{7}{24}\delta+\Bigl(\frac{11}{20}\eta-\frac{1}{8}\xi\Bigr)\frac{\pn}{2M}\biggr].
\end{equation}
Thus, for soft and hard photon emission in the cases we are interested in~(c.f. discussion in the introduction section,~\ref{sec:intro}), the mean lifetime $\tau$ of the electron can be expressed, as usual, in terms of the decay width~(\ref{eq:cdr2}) via
\begin{equation}\label{eq:elt}
\tau(\pv)\equiv 1/\Gamma_{e\rightarrow e\g}(\pv).
\end{equation}
For $\xi=0$ and $0<\eta$~(or $\eta=0$, $0>\xi$) of its generic order $\O(1)$ with $\delta\approx 0$, $\tau\sim 8\alpha_{e}^{-1}M/p^{2}$ and we can see that the difference in the effective radiation time $\tau$ between these two cases is only up to an $\O(1)$ numerical factor. While our primary focus is on the linearly Planck-scale suppressed modifications to the dispersion relations~(of the type arising from certain string inspired scenarios discussed in section~\ref{sec:sstf}), we note that qualitatively similar features have been observed for QED with dimension-4 LV modifications~\cite{Altschul:2007tn,Klinkhamer:2008ky}: as demonstrated through detailed analytical calculations in ref.~\cite{Altschul:2007tn}, the energy loss rates from the soft and hard parts of the Cherenkov photon spectrum are comparable in magnitude in this theory. It should be stressed that the $\O(E/M)$ dispersion modifications correspond to dimension-5 terms~\cite{Myers:2003fd} within the framework of effective field theories, though our setting~(\ref{eq:rdr}) further assumes helicity/spin independence.

Consider ultrahigh-energy electrons, for which one takes the highest energy~(indirectly) detected by LHAASO detectors~\cite{LHAASO:2021cbz}, $E\sim 2$~PeV. As mentioned, they are widely regarded as the origin of the observed UHE IC photons of the Crab Nebula. From eq.~(\ref{eq:elt}), therefore we have,
\begin{equation}\label{eq:elte}
\tau\propto10^{-11}\Bigl(\frac{10~\textrm{TeV}}{p}\Bigr)^{2}\times\O(1)~\textrm{s},
\end{equation}
where 10~TeV is the order of the threshold for $\O(1)$ $\eta$~(or $\xi$) as required by eqs.~(\ref{eq:sct}),~(\ref{eq:hct}). Then for $p=\O(\textrm{PeV})$, $\tau$ is just $\sim 10^{-15}$ seconds, implying that the initial electron/positron loses a significant fraction of its excess energy really quickly, corresponding to propagation over a distance scale of micrometers. Since it certainly travels distances far larger than that, we must demand that the threshold for the vacuum Cherenkov effect is above the maximal energy of any charged lepton known to be stable, regardless of whether the emitted photon is soft or hard. This is precisely the condition utilized for setting bounds upon LV via such effects, thus constraining $\xi$ or $\eta$ to at least of $\O(10^{-6})$~\cite{He:2022jdl,Li:2022ugz}. If one adopts such improved constraints, the decay rate will be $10^{6}$ times smaller and the particle radiates `all' its energy down to thresholds in a slightly longer time of about 100 picoseconds, which are still short enough, that on any relevant astrophysical timescales electrons of PeV energies will almost completely decay any time it takes place.

We therefore conclude that for a general class of Lorentz-violating dispersion relations, of the type~(\ref{eq:gdr}), in all cases of interest the ensuing electron Cherenkov process \emph{in vacuo} is an efficient reaction, once kinematically allowed, as justified by the rapidity with which the decay rate becomes large above thresholds. It may be considered for all practical purposes, as happening instantaneously and the loss rate $\dot{E}\sim e^{2}E^{3}/M$ is high enough that one does not make any relevant mistakes if just exploiting the absence of thresholds for the purpose of limiting certain types of scales of the violation. Hence in particular, the stringent bound imposed by simply looking at the threshold condition~(of the hard Cherenkov effect) on the linear LV scale entailed by a modified subluminal photon dispersion, $E_{\textrm{LV}\g}=M/(-\xi)^{-1}\gtrsim 1.2\times 10^{25}$~GeV~(if $\eta=0$)~\cite{He:2022jdl} is indeed reliable at least as long as we work in the realm of the purely phenomenological LV extension to the canonical relativistic dispersions. But, as we shall see in the next section~(section~\ref{sec:sstf}) there exist QG models where such a constraint does not apply and where there would be compatibility~\cite{Li:2021gah} with the hint of light-speed variation from time delay measurements of GRBs~\cite{Shao:2009bv} that require a value of order $10^{17}$~GeV for the same scale parameter.

\subsection{\label{sec:com}Comparison with analogous models of QED}

Before closing this section, we would like to compare our result for radiation rates with the ones in some selected models, which also allow the emergence of vacuum Cherenkov effect. In ref.~\cite{Klinkhamer:2008ky}, the following expression for the total power radiated by the charge is obtained, for modified Maxwell theory:
\begin{equation}\label{eq:elrm}
P(E)\equiv -\frac{\sd E}{\sd t}\approx\frac{e^{2}}{4\pi}E^{2}\biggl[\Bigl(\frac{7}{24}\kappa-\frac{1}{16}\kappa^{2}+\O(\kappa^{3})\Bigr)+m_{e}\textrm{-dependent terms}\biggr],
\end{equation}
where, as far as the form of the resulting deformed dispersions is concerned, the role of the LV parameter, $\kappa\equiv 2\tilde{\kappa}_{\textrm{tr}}(>0)$ for this theory may be played by $\delta$ here. The energy behavior of the decay rate, in the leading order in $\kappa$, is clarified by
\begin{equation}\label{eq:cdrm}
\Gamma_{e\rightarrow e\g}=(2\kappa/3)E\alpha_{e}.
\end{equation}
As we expect, in the case $\eta=\xi=0$, the leading-order corrections to rates, estimated from our phase space analysis~(\ref{eq:elr}),~(\ref{eq:cdr2}) agree then with those of~\cite{Klinkhamer:2008ky}. On the other note, as for the terms of linear Planck-mass suppression~(more relevant for us), the radiated-energy rate found from~\cite{Jacobson:2005bg}, in the context of effective field theory approach to analogous QED~\cite{Myers:2003fd} with the broken CPT, has similar behaviors just as our result indicates: the net decrease in the charged-lepton energy above threshold is controlled by $\lvert\sd E/\sd t\rvert\sim p^{3}/M$~\cite{Mattingly:2005re,Jacobson:2005bg}, which again aligns with eq.~(\ref{eq:elr}). It is this fact that was previously used, within that framework, for justifying the adoption of kinematical analyses for phenomenological explorations of LV with vacuum Cherenkov effect, as we are tempted to argue here. Still, we further show that the result stays the same for a generic class of LV in-vacuo dispersions.~(Similar observation is recently pointed out also in~\cite{Artola:2024mky} while the case discussed is that of a complex scalar field.) Therefore, being adopted by many theory-independent LV analyses using astroparticles, the modified dispersion of eq.~(\ref{eq:gdr}), or its variant, by itself permits a fairly sensible estimate for the rate from~(kinematic) phase-space consideration, thus offering an additional reason for our ignorance of any possible modifications to the matrix element~(an aspect of dynamics) which is not quite important in this problem.

We make an important remark here.\footnote{We are grateful to an anonymous referee for bringing this issue to our attention and encouraging us to clarify this point for the reader's benefit.} The phase space estimates, which form the basis of our analysis, may not be entirely accurate for determining the rates of Lorentz-violating processes when modifications of particle dispersion relations depend on helicity and particle versus antiparticle identity. This is particularly true for dimension-5 LV QED~\cite{Myers:2003fd}: as shown in~\cite{Jacobson:2005bg}, properly accounting for this dependence complicates matters, generally altering the behavior of the decay rates. For a detailed discussion of the role that spin structure plays in this theory, we refer the reader to ref.~\cite{Altschul:2010nf}. However, as previously emphasized, our explicit calculations here are restricted to spin-independent LV cases. While such modifications are not allowed for photons in local field theory, they do align with the D-brane/string models presented later on.

Finally we mention for completion, that although the procedure outlined above leading to the results of eqs.~(\ref{eq:cdr2}),~(\ref{eq:elr}) is general, it pertains strictly speaking to energy regimes where mass terms can be simply ignored, i.e., well above threshold. Yet, masses are important for Cherenkov radiation, as they determine the energy thresholds, c.f.~(\ref{eq:sct}), and~(\ref{eq:hct}). However, in the case that initial leptons are near or only slightly above the threshold, their lifetimes are also short enough~($\ll 1$~s). To convince oneself of this it suffices to repeat the derivation following the method described in appendices of~\cite{Jacobson:2005bg} for photon decay. Since, in such a case, an electron, just above threshold, presumably would continuously radiate only low-energy lights, reaching the threshold value asymptotically, the process will be soft. We are therefore not very interested in computing this given that the inclusion of hard process is necessary for our discussion.

\section{\label{sec:sstf}Electron propagation in the space-time D-foam dispersive medium}

In the previous section we observed for general LV dispersions~(\ref{eq:gdr}), that the electron decay once permitted by kinematics, is so efficient that one can neglect the actual rate and derive threshold constraints, via a direct comparison with the LHAASO data of the Crab Nebula. Such results may be translated also into some of the strongest constraints on~(field theory) LV models mentioned in subsection~\ref{sec:com}, as already noted in our ref.~\cite{Li:2022ugz}. However, although the lack of a detectable \emph{hard} Cherenkov effect allowed us to even push the scale, $E_{\textrm{LV}\g}$ that characterizes a \emph{subluminal} LV \emph{dispersion} relation of the \emph{photon} up to six orders higher than the Planck scale it does \emph{not} exclude all possible such modifications of photon propagation. Indeed, as we shall point out~(in section~\ref{sec:evs}), in some brane-world/string models for quantum space-time foam situations~\cite{Ellis:1999uh,Ellis:2000sf,Ellis:2004ay,Mavromatos:2005bu,Ellis:2008gg,Ellis:2009vq,Li:2021gah,Li:2023wlo,Li:2024crc} analyzed in this section, one cannot use them to cast similarly tight limits on the model parameters. Such scenarios hold subluminous \emph{refractive} indices of linear $\Mp$ suppression, only for neutral~(gauge) particles~\cite{Ellis:2003sd,Ellis:2003ua} like photons, but not for electrons so that the latter could travel faster than photons of comparable energies. Naively, this may lead to~(hard) Cherenkov effects, however additional mechanisms at work preclude that possibility. Subsection~\ref{sec:ace} shows examples of how this works in certain brane models provided by the approach.

\subsection{\label{sec:dfm}String/Brane-world inspired D(efect)-foam and electrons}

Below, we hope to present the plausible mechanisms that could permit stable propagation of electrons for this type of QG-foam of strings. Actually, for any theory exhibiting subluminal refraction effects of photons in a nontrivial vacuum, we have emphasized in the introduction that, one can avoid the occurrence of the process, $e\rightarrow e\g$ \emph{in vacuo}, if the theory entails the \emph{lack} of \textbf{i)} a description of the photon propagation via the dispersion relation, \emph{or} \textbf{ii)} an exact energy--momentum conservation. String models we shall consider below are characterized by these two unusual~(and useful) properties, respectively, related with their lack of a local-effective Lagrangian formalism. We must stress at this point, that an oversimplified version of the suggestion that electrons do not Cherenkov radiate in the foam, was first put forward in~\cite{Ellis:2003ua}, but to our knowledge, no detailed description has been presented, especially for the models developed subsequently, i.e., in~\cite{Ellis:2008gg,Ellis:2009vq} and~\cite{Li:2023wlo}, two specific situations that we are concerned with, thereby calling for some clarifications.

For completeness, let us first review briefly the basics of the approach as well as its phenomenology. The inclusion of D-branes, $D$-dimensional string solitons, in this theory leads to a scattering picture of microscopic QG fluctuations, i.e., \emph{space-time foam}~\cite{Wheeler:1998stf}. Typically, a D0-brane~(D-particle) is allowed in certain string theories like type IIA and I$^{\prime}$~(or IA) but here we consider it to be present in models of phenomenological interest~\cite{Mavromatos:2005bu,Ellis:2008gg,Ellis:2009vq,Li:2021gah,Li:2023wlo,Li:2024crc} since, even when 0-branes do not exist, as in IIB theory~\cite{Li:2009tt}, there are effectively pointlike D-`particles' from wrapping up D3-branes around 3-cycles. When low-energy matter, viewed as an open string propagating in a $(D+1)$-dimensional space-time is captured and later re-emitted by a defect embedded in this space-time, the recoil of the D-particle~\cite{Ellis:1999uh} and its back-reaction will modify the string propagation~\cite{Ellis:1996lrt,Kogan:1996zv,Ellis:1997cs}.

These considerations lead to a supersymmetric construction for space-time foam which could have realistic cosmological properties~\cite{Ellis:2004ay} in a brane-world scenario~\cite{Maartens:2003tw}. It is based on parallel brane-worlds where the standard model particles are anchored, moving in a higher-dimensional bulk space which contains a `gas' of I/IIA D-particles~(0-branes). One of these branes, after appropriate compactification to three spatial dimensions, represents allegedly our observable Universe. As it sweeps through the bulk, D-particles traverse it in a \emph{random} way. From the point of view of a brane observer, the crossing D-particles look like twinkling background \emph{defects}~(hence the name D(efect)-foam), giving 3D brane~(our world) a \emph{D-foamy} structure. The propagation of photons is affected, as originally suggested within noncritical string theories~\cite{Amelino-Camelia:1997ieq}, due to their interactions with such defects. Although till today this has not been handled analytically as a second-quantized formalism for string theory is lacking, the space-time distortion induced by the intermediate composite string excited in quantum scattering off the D-particle may be calculated perturbatively, for nonrelativistic defects~(of mass $\Md=\Ms/\gs$, with $\gs$ being the string coupling and, $\Ms$, the string mass), i.e., in the weakly-coupled string limit $\gs<1$. Schematically, therefore, the identification of the target time $t$ with the so-called Liouville mode, needed for restoring the lost conformal invariance for the $\sigma$-model which describes the open-string excitations corresponding to photon fields in first-quantized strings and that has been driven \emph{supercritical} by D-brane recoil, fixes the form of such a \emph{local} disturbance.\footnote{To aid the reader who consults the original reference, which essentially our analysis is based upon, some details are relegated to appendix~\ref{app}. In this subsection we just summarize previous main results.} For large $t$ it asymptotes to a metric, which requires the machinery of logarithmic conformal field theory~(LCFT)~\cite{Mavromatos:1998nz}:
\begin{equation}\label{eq:std}
G_{ij}=\delta_{ij},\quad G_{00}=-1,\quad G_{0i}\simeq u_{i}=\Delta k_{i}/\Md,
\end{equation}
where the defect's recoil velocity $u_{i}$ is small, $\lvert u_{i}\rvert\ll 1$, and its direction breaks target-space Lorentz invariance. Naively, the modified dispersion would be implied, for a photon, due to the metric~(\ref{eq:std}) which depends on \emph{both} the coordinate and momentum transfer, $\Delta\kv$, of the photon during its interaction with the defect and, so has similarities to a Finsler metric~\cite{Bao:2000rfg} whilst acquiring \emph{genuine} effects of the foam, viewed as a whole, from the macroscopic point of view of a 4D observer, is a delicate one, depending on the details of the \emph{average}~(denoted below by $\sdla\cdots\sdra$) over a defect ensemble, that is an issue that, in principle, may be handled differently. Nonetheless models of our interest~\cite{Ellis:2008gg,Ellis:2009vq,Li:2023wlo} agree in predicting time delays which are proportional to the frequency, $\wa$, for the more energetic photons,\footnote{For the sake of clarity, notations like $\pa_{i}$, $\wa$, and, $\Ea$, are sometimes used to imply the expectation values in the considered foam vacuum $\sdla\cdots\sdra$'s, that are identified as observed quantities.}
\begin{equation}\label{eq:dfptd}
\Delta t^{\textrm{D-foam}}\propto\wa/\Ms,
\end{equation}
provided a refractive index $\eta_{D}$ of the respective foam vacuum. We shall see however that its origin could be brane model dependent: in the stretched-string foam model it enjoys~\cite{Ellis:2008gg,Ellis:2009vq} a nonperturbative description while it arises from metric of the type~(\ref{eq:std}) but of a \emph{stochastic} kind in models considered in~\cite{Li:2023wlo}. Although D-foam also has effects in cosmology leading to supercritical string cosmologies or~(dissipative) Liouville $Q$-cosmologies~(with $Q$ the central charge deficit) no significant constraint was imposed~\cite{Mavromatos:2010pk}.

Comparison with the evidence for delayed arrivals of multi-GeV cosmic $\g$-rays of GRBs from Fermi~\cite{Fermi-LAT:2009grb} implies useful constraints on the density of defect populations on our brane. Indeed, although the mechanism of radiation of GRB still eludes us, a series of analyses~\cite{Shao:2009bv} combining data from multiple sources reveal a significant correlation between time lags and particle energies, $\Delta v_{\g}=-E_{\g}/E_{\textrm{LV}\g}$, with a~(photon) LV scale roughly compatible with the order of Planck scale~($E_{\textrm{LV}\g}\sim 3\times 10^{17}$~GeV), implying a \emph{subluminal} light-speed variation, which could be attributed to D-foam induced refractive indices \emph{in vacuo}, as we have argued previously~\cite{Li:2021gah,Li:2021tcw}, and which, upon introducing as a hypothetical situation that for redshifts $z\lesssim 10$ the foam is \emph{uniform}, amounts~\cite{Li:2021gah,Li:2023wlo} to constraining quantities like:
\begin{equation}\label{eq:cdf}
\frac{\Ms}{\gs\nd(z)}\geq 3\times 10^{17}~\textrm{GeV},\quad \nd(z<10)\simeq\textrm{const.},
\end{equation}
which plays the role of the QG scale\footnote{Note that from a modern perspective, where large extra dimensions may be introduced, $\Ms$ is considered arbitrary. Despite the discrepancies in the bounds placed on $E_{\textrm{LV}\g}$ like that provided by GRB~090510~\cite{Fermi-LAT:2009grb,Xiao:2009xe} where \emph{ad hoc} assumptions on the intrinsic time lags at the source have been made, it could indeed be probed with the findings from these analyses~\cite{Shao:2009bv}. The negative source timing~(i.e., high energy photons are emitted before softer ones) found solely via data fitting agrees with the cooling nature of GRBs. A recent survey~\cite{Liu:2024qbt} of the brightest ever GRB~221009A data finds direct signals for this \emph{preburst} stage of GRBs.} associated with frequency dispersion for the propagation speed of light, with $\nd$ being the D-particle number density, i.e., the number of defects encountered by the photonic matter per string length $\ls(\equiv\Ms^{-1})$. Such an explanation appears remarkably consistent also with the TeV spectra of active galactic nuclei~(AGNs)~\cite{Li:2020uef} which constitute some other examples of observed effects on retarded photon arrivals, with higher energy photons arriving later. It is also important to note that, the delay effect~(\ref{eq:dfptd}), being independent of photon helicity, is \emph{nonbirefringent}, evading the otherwise tight bounds from polarized measurements~\cite{Li:2025stf,He:2022gyk}~(see also ref.~\cite{Li:2021gah} in this regard).

After these necessary digressions concerning effects on photon propagation we now look into the case of electrons, as it is a prerequisite for exploring the fact, in the next subsection, that no electron decays can ever happen. The important point to notice, is that such string models for space-time fluctuations offer~\cite{Ellis:2003ua,Ellis:2003sd} an extreme version of \emph{equivalence-principle violation} in QG: D-particles have no `hair' with only vacuum quantum numbers, so they are neutral and can `cut' only photon open strings or at most electrically neutral particles~(say, neutrinos) which split into intermediate string states as a result. To charged ones, the foam looks \emph{transparent} since such a cutting procedure is forbidden, due to charge conservation.\footnote{It is strictly speaking the case only in I/IIA string foam where the defects are truly pointlike, and which we make use here. For IIB strings~\cite{Li:2009tt}, the~(wrapped-up) brane defects permit charge flow on their surfaces, so there are nontrivial, but much more \emph{suppressed} fermion--foam couplings.} Electrons are thus not supposed to induce the capture/recoil processes and we expect their canonical in-vacuo dispersions to be applicable~\cite{Li:2022ugz},
\begin{equation}\label{eq:dfedr}
\sdla p^{\mu}G_{\mu\nu}p^{\nu}\sdra\simeq -\Ea^{2}+\overline{\pv}^{2}=-m_{e}^{2}.
\end{equation}
This is consistent with a local effective action proposal for the model~\cite{Mavromatos:2010ar} in which charged fermion $\psi$ has no tree-level coupling to the foam recoil-velocity field $u_{i}$:
\begin{equation}\label{eq:dfea}
S_{\psi}=\int\sd^{4}x(-\bar{\psi}\slashed{D}\psi),\quad D_{\mu}=\partial_{\mu}-eA_{\mu},
\end{equation}
where $m_{e}$ is absent, since, at least in stochastic D-foam of~\cite{Mavromatos:2010ar}, the mass, being dynamically generated, is \emph{catalyzed} by the relevant LV photon-$A_{\mu}$-field terms, which capture part of the foam effects except the induced delays.

\subsection{\label{sec:ace}Absence of \emph{any} Cherenkov effect}

Let us now justify the absence of electron Cherenkov effect, in this framework, adding some details to previously published work~\cite{Li:2022ugz}. Without further detailed analysis all we did, there, is just reiterating this conjecture based solely on the fact that~\cite{Ellis:2003ua} electrons are not affected by interactions with D-foam vacuum medium, i.e.,~(\ref{eq:dfedr}). By itself this is insufficient, since, the naive correspondence of the present model(s) to the LV parameter configuration $(\eta=0$, $\xi<0)$ for the generic phenomenological approach to nonstandard particle dispersions~(\ref{eq:gdr}) could indeed allow one to imply a hard emission process. However as a result of mechanisms exclusive to concrete models of foam electrons are stable~(practically at least) to this effect. We examine two such examples below.

\subsubsection*{The stretched string formalism}

In~\cite{Ellis:2008gg}, a nonperturbative mechanism for the induced time delays~(\ref{eq:dfptd}) has been developed, based on a microphysical description of the capture process of the stringy matter by a defect within standard \emph{critical} string framework. In particular, there are analogies of the situation with the familiar local-field-theory description of a refractive index for wave propagation in conventional solid media where the role of electrons, treated as simple-harmonic oscillators, is played here by D-particles. Contrary to solid-state case, in the string model we now come to discuss, the corresponding `gravitational medium effect' does not admit an effective field-theory formulation and is found proportional to the photon energy, of the kind reported in the intriguing analyses~\cite{Shao:2009bv,Li:2020uef} of flight-time lags of cosmic photons.

The main advantage of this scenario is having a microscopic picture which is essentially \emph{stringy}, according to which as a photon strikes a brane defect there is an intermediate string state created, \emph{stretched} between the D-particle and the brane-world. This stretched string stores the energy $k^{0}(\equiv\omega)$ of the photon state as potential energy, by stretching to a length $\ell$ and acquiring some $N$ number of internal oscillatory excitations, c.f. the electron oscillators in the case of a solid material:
\begin{equation}\label{eq:pe}
k^{0}=\frac{\ell}{\alp}+\frac{N}{\ell},
\end{equation}
here $\alp=\ls^{2}$ is the Regge slope. The restoring force in the familiar atomic case, necessary to keep the electron quantum-oscillating about its position, is provided, here, by the stretched string, which carries characteristic flux of the D-brane interactions. The time taken for this string to grow to the maximal length $\ell_{\max}$, determined by the energy minimization~($L_{\max}=\alp p^{0}/2$), and then shrink back to zero~(i.e., decay) is~\cite{Ellis:2008gg},
\begin{equation}\label{eq:sstd}
\Delta t\sim\omega/\Ms^{2},
\end{equation}
This dictates then the time delay for a photon emerging after capture by a \emph{single} D-particle. Using the usual relation $\Delta t=(\eta_{D}-1)L$, in case of a foam~(with a linear density of defects $\nd/\ls$, on the brane) where the photon traverses a distance $L$, with its energy $\wa$ conserved as a mean-field approximation, one obtains a refractive index $\eta_{D}$, which indicates nontrivial `optical' properties for the brane, responsible for the final delay effect~(\ref{eq:dfptd}) experienced by the photon~\cite{Ellis:2010he,Li:2024crc}:
\begin{equation}\label{eq:sspri}
0<\eta_{D}(\wa)-1\simeq\nd\alp\wa\sigma_{\g}\Bigl[\frac{\gs}{\ls}\Bigl]\ll1,
\end{equation}
where $\sigma=\sigma[\ls^{-1}\gs]$ is the photon/D(-particle)-defect scattering cross section. Here we keep the dimension of $\sigma$ explicit as in our ref.~\cite{Li:2024crc} in order to make manifest the minimal~(linear) suppression of the effect by the string scale.

Some comments are in order here concerning the novel features~(that are of importance for our arguments on electron stability) for this model, in comparison to Liouville approach to recoil, outlined in section~\ref{sec:dfm}.
\begin{itemize}
\item The characteristic delays in this picture are due to topologically nontrivial collisions of strings with defects involving the creation of intermediate string~(and hence nonlocal) states. They are purely stringy effects associated with a time--space \emph{string uncertainty principle} $\Delta t\Delta X\geq\alp$~\cite{Yoneya:2000bt}, by making use of Heisenberg uncertainties, $\Delta X\geq 1/\Delta p>\alp/(\alp k^{0})\sim\alp/\Delta t$, where the momentum uncertainty $\Delta p<k^{0}$.
\item There is a formal analogy of the situation with that~\cite{Seiberg:1999vs} of strings in an antisymmetric $B_{\mu\nu}$-background of `electric' form~(i.e., $B_{0i}\neq 0$, $B_{ij}=0$), whose role is played here by D-particle recoil velocity $B_{0i}\sim u_{i}$. The induced space-time noncommutativity~\cite{Amelino-Camelia:1997pfm,Mavromatos:1998nz} corrects the above eq.~(\ref{eq:sstd}) by higher order terms: $\Delta t\sim\alp k^{0}/(1-\un^{2})$. Such delays occur in a way reminiscent of open-string/open-string scattering as calculated in~\cite{Seiberg:1999vs}, respecting \emph{causality}, in contrast to~(noncommutative) field-theoretic treatments, that suffer from noncausal effects.
\item These effects are inherently stringy/quantum gravitational and \emph{cannot} be interpreted, via some effective local low-energy Lagrangian, as an average propagation of photons. In fact, the uncertainty-caused delay~(\ref{eq:sstd}) holds even if average effects of recoil vanish, $\sdla u_{i}\sdra=0$, and only multi-point correlation exists, say, $\sdla u_{i}u^{i}\sdra\neq 0$. The associated anomalous photon dispersion terms in such a case induced by the Finsler metric~(\ref{eq:std}), depend quadratically on the particle energy, implying~\cite{Ellis:2010he} the \emph{disentanglement} of the linear QG effect~(\ref{eq:sspri}) from any possible recoil-induced modified dispersion which may find a local effective model representation.
\end{itemize}

It becomes clear now that the stringy uncertainty delay mechanism which characterizes exclusively the D-particle model of~\cite{Ellis:2008gg}, does \emph{not} entail any modification of the local~(microscopic) dispersion relations of photons. This is a crucial difference from those generic~(phenomenological and/or field-theory) LV settings analyzed previously, where anomalous photon propagation is expected via corresponding modifications of particle dispersions, thereby distinguishing this scenario from the latter ones. This has important consequences for phenomenology, in particular concerning the severe restrictions due to hard emission of vacuum Cherenkov radiation, as have to be imposed in the latter cases. However, these do not apply to the present model, for reasons we argue as follows.

To understand this point, one should first recall that these constraints were established based on analyses on a linearly deformed~(photon) dispersion relation, such as that of~(\ref{eq:rdr}). Regardless if it can be obtained from an effective action, like that of a flat-space theory~\cite{Myers:2003fd} involving higher-dimension local operators, or not, c.f. being just purely phenomenological, strict energy--momentum conservation is assumed in any interaction process. Although this should be re-examined in a Lorentz-violating QG framework, as stated before and has been pointed out also in~\cite{Ellis:2000sf,Li:2023wlo} in which the scenarios envisaged~(discussed later on) do entail a violation of this assumption, at the present level of ignorance we have toward this problem, in principle energy--momentum should, at least formally, be conserved for all practical~(low-energy phenomenology) purposes.

Indeed, within the formalism~(\ref{eq:pe})--(\ref{eq:sspri}) of string formation, stretching and decay that we now consider, there is no mechanism for momentum--energy nonconservation in 3-point vertices, precisely because the above stringy~(microscopic) method developed for describing the capture and recoil process is formulated with the language of critical string theory. Also, in this context, for excitations that appear as external states in amplitudes, the appropriate kinematics is that these states should all be regarded as on mass-shell, for reasons we have explained. This is to be contrasted with models based on metric distortion~(\ref{eq:std})~(see below) where energy fluctuations appear in particle interactions, despite momentum conservation, and where that metric applies to the external photons. Nonetheless even in the latter case, one is still faced with a refractive index of the form~(\ref{eq:sspri}) but, as we discuss below, it is due to specific aspects of the model regarding photon--foam interaction.

Now that in the formulation of~\cite{Ellis:2008gg}, the process $e\rightarrow e\g$ \emph{in vacuo} involves only external photon and electrons, and Lorentz violation is absent for electrons, and, also for the photon but at the level of the dispersion-relation, this effect is forbidden by the energy--momentum conservation, as in the usual Lorentz invariant theory. The current observational null result of the effect therefore, does \emph{not} constrain this framework. Possible effects due to nontrivial higher-order recoil contributions, which, for the most physically intriguing case of vanishing average recoil $\sdla u_{i}\sdra=0$ we are interested in, might induce terms of the form 
\begin{equation}\label{eq:sspdr1}
\wa^{2}-\overline{\kv}^{2}\sim\O\Bigl(\overline{\kv}^{2}\gs^{2}\frac{\wa^{2}}{\Ms^{2}}\Bigr),
\end{equation}
need to be investigated but it is clear that the induced dispersion corrections are suppressed by higher powers of $\Ms$~(to lowest order, the square of the D-particle mass scale as in~(\ref{eq:sspdr1})) if they appear at all. Then, for $\lvert\kv\omega\rvert\ll (\Ms/\gs)^{2}$, as appropriate for a photon of astrophysical origin when assuming traditionally high string scales~(say, $\Ms\gtrsim 10^{16}$~GeV), eq.~(\ref{eq:sspdr1}) looks roughly like a standard relativistic dispersion $\wa\sim\ka$, suggesting again the \emph{intactness} of the kinematics for $e\rightarrow e\g$~(i.e., \emph{forbidden} as usual). We remark that in a T-dual picture, where metric~(\ref{eq:std}) due to recoil becomes diagonal~(c.f. eq.~(\ref{eq:app8}) of appendix~\ref{app}) this appears as a rigorous result in the limit of small recoil velocities $\lvert u_{i}\rvert\ll 1$ where we restrict our attention throughout the work:
\begin{equation}\label{eq:sspdr2}
\wa^{2}=\overline{\kv}^{2}+\Bdla\frac{m_{\g}}{1-\un^{2}}\Bdra\biggl\vert_{m_{\g}=0}=\ka^{2},
\end{equation}
where the masslessness of the photon is disentangled in the model from the linear-in-energy time delay~(\ref{eq:sstd})~(and hence the induced nontrivial vacuum refractive index effect), realized through string/brane uncertainties induced by the stretched string between D-brane defects of the foam and the brane-world.

In fact, analogous arguments were previously provided to reconcile with pertinent facts regarding testing the predictions of the model via neutrino astronomy in ref.~\cite{Li:2024crc}, where we indicated that several neutrino-related channels remain conventional as well. For electron's behavior that is of interest to us here, partially, the same subject has been discussed in~\cite{Mavromatos:2010pk}, but with a focus on facts that the formalism avoids modifications to the Crab synchrotron radiation~(and thus the resulting constraints from modifying electron dispersion relations). As we hope becomes clear from our discussions which complete the studies in refs.~\cite{Ellis:2008gg,Mavromatos:2010pk}, Cherenkov effect is also not allowed since UHE electrons~(positrons) obey the usual Lorentz kinematics, despite having greater velocities than that of the photons which experience QG refraction~(\ref{eq:sspri}). As we have reiterated, the latter effect is purely stringy, not represented by the above Finsler dispersions governing particle interactions, and so it cannot be restricted by considerations on vacuum Cherenkov effect.

\subsubsection*{Revisit a random isotropic foam}

For our brane-world, as we mentioned, D-particle crossings constitute a realization of foam. Their effect is actually `classical' from the bulk space-time viewpoint, but appears quantum-mechanical for a low-energy brane observer. Although, in this sense, it is possible to propose some kind of effective model for D-foam, for reasons stated, it is not a simple task to do this. In~\cite{Li:2023wlo} for example, we nevertheless show how, in principle, such a model can be formulated in order to get consistency with certain phenomenological hints of LV for cosmic neutrinos. Recoil fluctuations \emph{along} our brane are considered in such cases a stochastic fraction of the incident momentum of the string state~\cite{Mavromatos:2005bu,Ellis:2004ay} so that $u_{i}$ in~(\ref{eq:std}) can be modeled as a local operator using the following parametrization:
\begin{equation}\label{eq:rvp}
u_{i\Vert}=\gs\frac{\zeta_{i\Vert}}{\Ms}k_{i},\ \textrm{no sum}~i\quad\Rightarrow\quad\sdla\zeta_{\Vert}\sdra=0,\ \sdla\zeta_{i\Vert}\zeta_{j\Vert}\sdra=\D_{i}^{2}\delta_{ij}\neq 0,
\end{equation}
which represents the effect averaged out by a brane observer. The suffix $\Vert$, appearing above, indicates components along brane longitudinal dimensions $i=1,2,3$ for the~(dimensionless) momentum fraction variable $\zeta_{i\Vert}=\Delta k_{i}/k_{i}$, which is taken to be Gaussian.

In general, the random foam implies vanishing correlators of odd powers of the velocity $u_{i}$, but nonzero correlators of even powers. The \emph{zero mean} of the distribution, characterized by~(\ref{eq:rvp}), restores Lorentz invariance \emph{on average}, whilst stochastic $u_{i}$ fluctuations break the symmetry in terms of the respective variance $\D_{i}$, and, cosmically, macroscopic effects of such a violation arise~\cite{Mavromatos:2010nk}. The form of this stochastic Finsler metric served as the starting point of our approach to Lorentz- and CPT-breaking neutrinos in~\cite{Li:2023wlo}, where kinematical aspects of matter/defect scattering are at play. The latter proves important for including plausibly unique propagation effect, i.e., subluminal neutrino~($\nu$) and superluminal antineutrino~($\bar{\nu}$), governed by the final dispersion relations:
\begin{equation}\label{eq:sfndr}
\Ea_{\nu}\simeq\Bigl(\lvert\overline{\E_{M}}\rvert/\pa-\frac{\pa}{2\Md}\Dn^{2}\Bigr)\pa,\quad \Ea_{\bar{\nu}}\simeq\Bigl(\lvert\overline{\E_{M}}\rvert/\pa+\frac{\pa}{2\Md}\Dn^{2}\Bigr)\pa,
\end{equation}
and, $\E_{M}\coloneqq\pm\sqrt{p^{i}p_{i}+m_{\nu}^{2}}$, where $m_{\nu}$ is the neutrino mass. We next considered an \emph{isotropic} foam such that $\sdla\zeta_{\Vert}^{2}\sdra\vert_{i=1,2,3}\equiv\Dn^{2}$, where the suffix $\nu$ implies that the variance $\Dn^{2}$ characterizes QG effects for neutrinos~(antineutrinos), but not other particles, due to nonuniversality of foam among particle species.

To study the issue concerning electron in-vacuo Cherenkov effect, we first derive photon dispersion relation, due to refraction effects of this foam. We shall be brief, since it parallels the case of neutrinos and the result is similar to~(\ref{eq:sfndr}). As we will explain, there is, however, a major difference, concerning photon subluminality. In the spirit of~\cite{Li:2023wlo}~(see appendix~\ref{app}), the back-reacted metric~(\ref{eq:std}) affects the energy--momentum relation for a massless particle, via $k^{\mu}k^{\nu}G_{\mu\nu}(k)=0$ for which the leading-order solution reads
\begin{equation}\label{eq:los}
\omega(k)\simeq \E_{M}+\frac{\gs}{\Ms}(k^{i})^{2}\zeta_{\Vert}+\frac{\E_{M}}{2}\frac{\gs^{2}}{\Ms^{2}}(k^{i})^{2}\zeta_{\Vert}^{2},
\end{equation}
where $\E_{M}\coloneqq\pm\sqrt{k^{i}k_{i}}$ now defines the Minkowski energy of \emph{indefinite} signature for photons. Taking its average over D-particle ensembles yields a quantity playing the role of the average energy for a photon moving through the foam:
\begin{equation}\label{eq:pae}
\wa\simeq\pm\Bigl(1+\frac{\gs^{2}}{2\Ms^{2}}\sideset{}{'}\sum_{i}\Dg^{2}k_{i}^{2}\Bigr)\times\overline{\kv},
\end{equation}
where $\sum_{i}^{\prime}(\Dg^{2}k^{i}k_{i})\eqqcolon(\Dg\overline{\kv})^{2}$ represents $\sim\sdla\zeta_{\Vert}^{2}(k^{i})^{2}\sdra$ schematically, due to~(\ref{eq:rvp}), and, the $\Dg$ is the corresponding variance. Meanwhile the kinematical aspects of recoil-defect/photon scattering enter the dispersion relation via 
\begin{equation}\label{eq:kae}
\wa=\wa^{\prime}+\Bdla\frac{1}{2}\Md u_{\Vert}^{2}\Bdra+\textrm{higher order terms},\quad u_{\Vert}\coloneqq\lvert\uv_{\Vert}\rvert,
\end{equation}
with the next order being $\O(u^{4})$. This therefore, implies,
\begin{equation}\label{eq:pdro}
\wa^{\prime}(\overline{\kv}{}^{\prime})\simeq\wa(\overline{\kv}{}^{\prime})-\frac{\gs}{2\Ms}\sideset{}{'}\sum_{i}\Dg^{2}(k_{i}^{\prime})^{2},
\end{equation}
where $\Ms u_{\Vert}^{2}/(2\gs)$ is the kinetic energy for a massive~(nonrelativistic) D-particle and $(\wa^{\prime},\overline{\kv}{}^{\prime})$ stands for the averaged 4-momentum of the re-emitted photon. In eq.~(\ref{eq:pdro}) the momentum conservation, c.f. $\overline{\kv}=\overline{\kv}{}^{\prime}+\overline{\Delta\kv}=\overline{\kv}{}^{\prime}+\Md\sdla\uv_{\Vert}\sdra=\overline{\kv}{}^{\prime}$ was utilized; this outgoing (average) energy suggests the total combined effects of foam on dispersion relations from both metric distortion and capture/splitting.

It is essential to recall that the difference between in-vacuo dispersions of neutrinos and antineutrinos~(\ref{eq:sfndr}) follows from Dirac's proposal of `hole theory', that previously served as a natural way to interpret the negative branch of eq.~(\ref{eq:pdro}): $\Ea_{\bar{\nu}}\equiv -\wa^{\prime}(\ka)>0$, for a physical antineutrino at high energies~($\Ea_{\bar{\nu}}\gg m_{\nu}$ so that $m_{\nu}$ is negligible). Such a consideration can apply only to fermions, as opposed to bosons, given that this `hole interpretation' argument is based upon the exclusion principle~\cite{Mavromatos:2012ii}. Hence, energetic photons are kept \emph{subluminal} for their bosonic property, exhibiting \emph{no} CPT violation, that will be with however for neutrino species, as they propagate in this medium of~(brany) vacuum defects. From eq.~(\ref{eq:pdro}) with $\wa^{\prime}$ positive, the associated LV dispersion can be read out as
\begin{equation}\label{eq:sfpdr}
\wa(\overline{\kv})\simeq\ka-\frac{\gs}{2\Ms}\Dg^{2}\ka^{2}=\ka\bigl[1-\O(\nd\wa/\Md)\bigr],
\end{equation}
to leading order in $\Dg^{2}$~(for $\Dg^{2}\ll 1$ which by construction, can be naturally up to $\O(1)$). The reduced propagation velocities for higher-energy photons, implied by~(\ref{eq:sfpdr}), from which an effective vacuum refractive index is expected,
\begin{equation}\label{eq:sfpri}
\eta_{D}=1+\O(\wa/\Ms),
\end{equation}
can then be tested experimentally, via searching for different arrival times for photons~(say, of GRBs) in different energy channels due to the existence of time delays, of the form~(\ref{eq:dfptd}), in particular concerning the light-speed variation proposed previously using cosmic photon data~\cite{Shao:2009bv,Li:2020uef}, c.f.~(\ref{eq:cdf})~\cite{Li:2023wlo}.~(We stress that despite being rather encouraging this latter observation as well as the strategy of analysis it relies upon remains preliminary and therefore should be treated with care.)

Next we come to the analysis on the~(hard) Cherenkov effect of electrons in the presence of foam. At this stage, we reiterate that the formalism of~\cite{Li:2023wlo} goes beyond the local effective Lagrangians that exact energy--momentum conservation in the process is, by its very nature, based upon, due to the presence of Finsler dynamical background~(\ref{eq:rvp}), where foam defects act as an external quantum-gravity `environment', resulting in \emph{energy violations} for particle reactions~\cite{Ellis:2000sf}. As we discussed in~\cite{Li:2023wlo}, for a $1\rightarrow 2$-body decay process~(relevant for us here), the induced energy nonconservation in the \emph{observable} particle subsystem, is associated with the kinetic energy of the heavy D-particle in the recoiling foam:\footnote{For a $\N$-particle interaction, a theorem holds, stating, that the total energy of all the observable particles \emph{is in general violated} by the requirement of \emph{energy conservation} for the complete matter $+$ recoiling D-brane system:
\begin{equation*}
\prescript{(\N)}{}{\sum\nolimits_{\ell}^{\pm}}E_{\ell}\simeq\frac{\Ms}{2\gs}u_{\Vert (\N)}^{2},\quad\prescript{(\N)}{}{\sum\nolimits_{\ell}^{\pm}}\pv_{\ell}=\Md\uv_{\Vert (\N)},\quad\Rightarrow\quad\delta\Ea_{D}^{(\N)}\coloneqq\overline{\prescript{(\N)}{}{\sum\nolimits_{\ell}^{\pm}}E_{\ell}},
\end{equation*}
here $u_{\Vert (\N)}$ is a proper generalization of~(\ref{eq:rvp}), c.f. $\uv_{\Vert}=\sum_{\ell}^{\pm}(\pv_{\ell}/\Md)$ and the $\sum^{\pm}$ represents, e.g., in~(\ref{eq:tev}): $\sum^{\pm}\! E_{\ell}=E_{\textrm{in}}-\sum E_{\textrm{out}}=E_{1}-E_{2}-E_{3}$ as $\N=3$. For details see appendix A of~\cite{Li:2023wlo} and/or ref.~\cite{Ellis:2000sf}.}
\begin{equation}\label{eq:tev}
\Ea_{1}-(\Ea_{2}+\Ea_{3})\eqqcolon\delta\Ea_{D}^{(3)}\simeq\frac{\Md}{2}\sdla\uv_{\Vert (3)}^{2}\sdra\neq 0,
\end{equation}
where $\Ea_{1}$ and $\Ea_{\ell}$~($\ell=2,3$), denote the~(average) energies for the incoming particle~(labelled 1) and outgoing particles~(2 and 3), respectively. A new dimensionless parameter, $\varsigma_{\I}$~(which $\neq\D_{I}^{2}$, in general), parametrizing the amount of such violations, is introduced via estimating the above quantity:
\begin{equation}\label{eq:tel}
\delta\Ea_{D}^{(3)}=\frac{\gs}{\Ms}(\varsigma_{\I}{}^{I})\pa_{2}^{2}+\ldots,\quad\varsigma_{\I}{}^{I}>0,
\end{equation}
where $\pa_{2}$ denotes typically the momentum of a particle emitted during decay, say, of particle 2, and particle 1 is converted to particle 3; the $\ldots$ on the r.h.s. denote contributions, for all practical purposes, negligible relative to the $\varsigma_{\I}{}^{I}$-term. We added the superfix $I$ in order to make manifest the particle-type dependence of the effect, $I=\g,\nu$. The fact that $\delta E_{D}^{(3)}\geq 0$, i.e., energy is \emph{lost} in particle decays with this random D-foam, follows from the underlying $\sigma$-model LCFT treatment~\cite{Mavromatos:1998nz,Kogan:1996zv} of recoil. As we shall explain, the presence of such losses does affect relevant energy-threshold equations, for the reactions to occur, which stem from kinematics, in such a way that certain otherwise allowed processes by generic LV, c.f.~(\ref{eq:gdr}), especially electron decays \emph{in vacuo}, are inhibited.

For this purpose, it suffices to calculate the threshold energy. We consider hard threshold, as it is the only effect which is allowed in the model, and with which subluminal Lorentz violations in cosmic photons may be limited. Let $\Ea$, $\pa$ be the~(average) energy and momentum of the incoming charged lepton, say of an UHE electron inside the Crab, with $(\Ea{}^{\prime},\overline{\pv}^{\prime})$, $(\wa,\overline{\kv})$ denoting the 4-momenta of the outgoing fermion and the emitted photon. The foam deformed conservation law is given by~(\ref{eq:tev}), that is,
\begin{equation}\label{eq:edcl}
\Ea=\Ea{}^{\prime}+\wa+\delta\Ea_{D}^{(3)},\quad\pa=\pa^{\prime}+\ka,
\end{equation}
where the spatial 3-momentum is conserved \emph{statistically}, as a result of the foam-background stochasticity condition~(\ref{eq:rvp}). The equations are already written in threshold configuration in which momentum is almost collinear. We exploit the fact that in D-particle foam models, Lorentz-noninvariant QG propagation effect is absent for charged excitations, as explained, so that $\Ea{}^{(\prime)}\simeq\pa^{(\prime)}\{1+m_{e}^{2}/[2\pa^{(\prime)2}]\}$~(see eq.~(\ref{eq:dfedr})), and that the hard emitted photon takes half of the incoming energy at threshold, to imply,
\begin{equation}\label{eq:thre}
-2m_{e}^{2}=E_{th}\Bigl(4\delta\Ea_{D}^{(3)}-\frac{\gs}{2\Ms}\Dg^{2}E_{th}^{2}\Bigr),\quad E_{th}\coloneqq E_{e,th}^{\textrm{D-foam}}
\end{equation}
where $\pa\simeq E_{th}=2\pa^{\prime}=2\ka$ and the amount of energy loss $\delta\Ea_{D}$ is essentially determined by eq.~(\ref{eq:tev}) with $\ka\simeq\pa_{2}=\pa/2$, to leading order:
\begin{equation}\label{eq:ela}
\delta\Ea_{D}^{(3)}\simeq\frac{\varsigma_{\I}{}^{\g}}{4\Md}E_{th}^{2}+\ldots,
\end{equation}
where according to~\cite{Li:2023wlo} we assume this equality to hold. The condition, that the relativistic electron/positron of momentum $\pv$, responsible for producing UHE IC radiation of the Crab Nebula, is near or below the corresponding threshold implied by~(\ref{eq:thre}),
\begin{equation}\label{eq:thrv1}
E_{th}\simeq\biggl[4m_{e}^{2}\frac{\Ms}{\gs(\Dg^{2}-2\varsigma_{\I}{}^{\g})}\biggr]^{1/3},
\end{equation}
amounts to the following inequality:
\begin{equation}\label{eq:thrc}
-\frac{\gs}{2\Ms}\bigl(2\varsigma_{\I}{}^{\g}-\Dg^{2}\bigr)\pn^{3}\leq 2m_{e}^{2}.
\end{equation}

The analysis so far we have performed has a few important outcomes that appear clear from eq.~(\ref{eq:thrv1}): this energy threshold exists only if $\sqrt{2\varsigma_{\I}{}^{\g}}<\Dg$ otherwise it would be either infinite~(for $\Dg^{2}=2\varsigma_{\I}{}^{\g}$) or meaningless~($\Dg^{2}<2\varsigma_{\I}{}^{\g}$), implying that hard Cherenkov emission is forbidden in a vacuum, with the special case of $\Dg=0$, $\varsigma_{\I}{}^{\g}=0$, as in the standard model. In case $\Dg^{2}>2\varsigma_{\I}{}^{\g}$, one can even envision the situation where this happens with $\varsigma_{\I}{}^{\g}\approx\Dg^{2}/2$, so the process could still be effectively inhibited. In this case,\footnote{Lacking, at present, a complete theory of matter--D-foam interactions we do not consider the rate above threshold. To first order in $1/\Md$, the matrix element for any process agrees with that of special-relativistic QED while recoil effects of D-particles modify the \emph{kinematics} of the field-theoretic result~\cite{Li:2023wlo}.} as expected, our arguments suggest a threshold that would be pushed to a very high energy scale of, for instance, order PeV or above, such that PeV electrons will not be depleted at all, and, can then propagate stably over distances required for subsequent IC scattering inside the nebulae. Thus we are able to observe $\g$-ray events of such UHEs as reported by LHAASO~\cite{LHAASO:2021cbz} in spite of implying a very tiny difference between $\varsigma_{\I}{}^{\g}$ and the half variance~(\ref{eq:rvp}):
\begin{equation}\label{eq:qgpc}
\lvert2\varsigma_{\I}{}^{\g}-\Dg^{2}\rvert\lesssim 8.6\times 10^{-26}\frac{\Ms}{\gs}~\textrm{GeV}^{-1}\quad\textrm{if}~\Dg^{2}\gtrsim 2\varsigma_{\I}{}^{\g}.
\end{equation}
This means that, exploiting the fact that these high-energy electrons/positrons do not hard Cherenkov radiate up to $\sim 2.3$~PeV, can only offer bounds for a \emph{combined quantity} of foam parameters, whilst the effective QG scale $\Md/\Dg^{2}$, associated with the~(subluminal) photon refraction of foam effective vacuum~(\ref{eq:sfpri}) is given by eq.~(\ref{eq:cdf}), $\Dg^{2}\sim 3\times 10^{-18}\Md~\textrm{GeV}^{-1}$. The constraint~(\ref{eq:qgpc}) can then be derived by putting $\Ms/\gs<3\times 10^{17}$~GeV as $\Dg^{2}\leq\O(1)$: $\Dg^{2}-2\varsigma_{\I}{}^{\g}\lesssim 3\times 10^{-8}$ where $\varsigma_{\I}{}^{\g}\sim 2\times 10^{-19}\Md~\textrm{GeV}^{-1}$.

We have to stress that these values imply \emph{no} extra suppression of the effect, but natural values for both parameters in such brane models where both $\Dg^{2}$ and $\varsigma_{\I}{}^{\g}$ being dimensionless though, are free to vary, independently. In case a lower~(TeV) $\Md$ is considered~(as in some low-scale string schemes~\cite{Antoniadis:1999rm}) the scenario envisaged by the data requires a smaller variance; this can also be naturally realized, with a much more dilute population of defects, provided $\Dg^{2}$ is proportional to the D-particle density, $\nd$, which is another free parameter of the bulk string cosmology model of~\cite{Ellis:2004ay} we used here. Although, from the observational requirement one concludes a limit~(\ref{eq:qgpc}) for $\Dg^{2}>2\varsigma_{\I}{}^{\g}$, this is not fine-tuning in our opinion, but implies the `recovery' of special-relativistic kinematics, which is indeed respected by strings, due to underlying Lorentz symmetry that is broken, or better, \emph{reduced} by the~(stringy) QG ground state. Indeed we have noted that the model may be characterized~\cite{Li:2023wlo} via a reduced-Lorentz symmetry~(akin to the one found from coupling~\cite{Ellis:1999sf} a fermionic system to the gravitational field~(\ref{eq:std})), with which the challenges brought about by superluminal neutrino instabilities, are properly resolved. These have not been discussed in our previous study of this isotropic stochastic D-foam, presented in~\cite{Li:2023wlo}, where attention was restricted to neutrino speed with favorable signal of the effect in astrophysical data.

For a soft emitted photon with an almost zero energy, however, its LV dispersion modification and the corresponding energy-violation effect during decay can be safely neglected. In this case, one can treat the process as it is in a field-theory based analysis, indicating the same requirement for the reaction to happen as shown in~(\ref{eq:sct}). However, the present model where neutral and not charged matter is susceptible to foam implies $\eta=0$; this means that this radiation effect is not permitted, proving the initial statement of~\cite{Li:2022ugz}, as also mentioned earlier. Hence, for the concrete class of string models examined here, anomalous Cherenkov effect \emph{in vacuo} is generically absent, at least up to the currently inferred maximum~(lepton) particle energy from the Crab Nebula.

While our analysis relies exclusively on the microscopic language of conservation laws, before closing this section, we briefly comment on its relationship to macroscopic approach. Classically, whether a moving electron radiates depends on whether the signal fronts of the radiation emitted at different points along its trajectory can interfere constructively in the far field. The speed that matters for the coherence condition ${\bf v}_{e}\cdot\kv=\omega(\kv)$~(which reduces in a LI vacuum with $\omega=kv_{ph}=k$ to $v_{e}\cos\theta_{C}=v_{ph}=1$, where $\theta_{C}$ is the Cherenkov angle) is the phase velocity $v_{ph}(k)=\omega/k$, determined by the dispersion relation. The alert reader should note that in the stretched-string model previously discussed, the apparent slowdown of photon propagation arises from nonlocal effects~(that is, global delays~(\ref{eq:dfptd})) which do not alter the phase matching for local interference. Since the photon dispersion relation remains unchanged in this theory, $\wa/\ka\neq 1/\eta_{D}(\wa)$, even if $v_{e}$ exceeds the effective average speed of photons delayed by D-particle interaction, $v_{e}<1$ still leads to destructive interference of the radiated waves, aligning with observations at the quantum level. Within models featuring a random foam, however, the phase speed of light may not be retained, potentially offering a classical explanation for vacuum Cherenkov effect under specific conditions. However, the analysis of the front velocity $v_{fr}=v_{ph}(\infty)$ requires a nonperturbative analysis going beyond the approximations made above for that case; this remains therefore an open question for further investigation.

\section{\label{sec:evs}Evasion of Cherenkov and other strict limits on photons}

We know that modified dispersions of the type~(\ref{eq:gdr}) if used to encapsulate QG phenomenology have a broad scope to limit LV. Besides vacuum Cherenkov effect focused on here, other effects regarding photon sector include time-of-flight delay in $\g$-ray detections. In~\cite{Shao:2009bv,Li:2020uef} in particular, it was suggested that, even with the incorporation of intrinsic timing of $\g$-rays at the respective source a tiny subluminal speed variation linearly scaling as photon energy could still be a significant contribution to the observed time delay, with an associated scale $E_{\textrm{LV}\g}\simeq 3\times 10^{17}$~GeV for the effect $\Delta v_{\g}=-E_{\g}/E_{\textrm{LV}\g}$. If one attempts to interpret it with the corresponding anomalous dispersion relations of the photons, the hard Cherenkov effect is allowed by kinematics as a result. It is in that context that UHE electrons deduced from the LHAASO measurements of the Crab IC spectrum have been claimed to raise the scale well above the Planck energy~\cite{He:2022jdl}; specifically, the presence of 2.3~PeV long living electrons requires the scale to be greater than $1.2\times 10^{25}$~GeV, which exceeds the sensitivity of GRB photon time delay studies to such subluminal photon LV properties~\cite{Shao:2009bv,Li:2020uef} by about seven orders of magnitude.

We also notice that, if one assumes modified dispersion relations for \emph{both} photons and electrons, as shown in~\cite{He:2022jdl}, the parameter space that is still permitted by the Crab 2.3~PeV electron jointly with the nonobservation of LV-induced $\g$-decay of the 1.4~PeV photon from a previous discovery of LHAASO~\cite{LHAASO:2021gok}, lies in a narrow region in the proximity of $\xi=\eta\leq 0$, which means that the LV scale of photons should be of similar order to the one of electrons if subluminal LV effects really act on them.\footnote{See fig.~3 of the second reference in~\cite{He:2022jdl}, where a joint constraint on photon--electron LV parameter space using also the photon decay, is presented.} However, as mentioned, and discussed in~\cite{Mavromatos:2010pk}, in order to avoid subluminal Lorentz-violating effects in the dispersion relation of electrons that would drastically affect~\cite{Jacobson:2002ye,Maccione:2007yc} the Crab synchrotron radiation that has been observed to extend at least up to energies of around 100~MeV~\cite{Atoyan:1996kqd}, just before the IC peak begins to contribute to the spectrum, by so-induced LV cutoffs of the synchrotron emissions that are not observed experimentally, one has to impose $E_{\textrm{LV}e}\gtrsim 10^{24}$~GeV~(which is derived at $2\sigma$). Then from the above-mentioned result of~\cite{He:2022jdl}, one might conclude that in the case of linear $\Mp$ suppression of the subluminal QG-motivated LV in photon propagation, of interest for the exotic~(QG) interpretation of light-speed variation effect~(\ref{eq:cdf})~\cite{Li:2021gah,Li:2021tcw}, scales with size $E_{\textrm{LV}\g}\simeq\O(E_{\textrm{LV}e})<10^{24}$~GeV are ruled out.

However, we stress that, this is not the case. The lack of comparison with a meaningful QG theory represents a severe limitation of the original analysis~(c.f.~\cite{He:2022jdl}) based on a purely phenomenological approach to LV dispersions. We point out that their result only \emph{excludes the possibility} that the intriguing findings of subluminal effects for cosmic photons~\cite{Shao:2009bv,Li:2020uef} are due to a LV effect that acts \emph{universally} among photons and electrons. Needless to say, it cannot exclude anomalous photon propagation with linear energy dependence in models where the Lorentz-violating QG foam is transparent to electrons which thus do not exhibit any nontrivial optical effects of the foam medium, as in the brane/string foam case~\cite{Ellis:2003ua,Li:2022ugz}. In view of our discussion in the last section, since no $e\rightarrow e\g$ processes can ever happen, at least practically, the requirement ensuing from an analysis relying on the putative electron decay that the subluminal light-speed variation can only exist at the similar level of the LV effect for electrons, dictated by the excluded regions in parameter space~\cite{He:2022jdl}, are simply not applicable, should the time delay of high-energy cosmic photons be due to light propagation in the above-described D-particle foam. In particular, the impressive~(and stringent) \emph{hard} Cherenkov bounds from~(\ref{eq:hct}) are naturally evaded, in two different ways depending on the explicit model considered.
\begin{itemize}
\item In the microscopic stretched-string model of D-particle foam as developed in refs.~\cite{Ellis:2008gg,Ellis:2010he} where the refractive index effect~(\ref{eq:sspri}) pertains to stringy uncertainties and as such is not linked to possible modified local dispersions, the kinematics and the interaction amplitude for $e\rightarrow e\g$ \emph{in vacuo} are largely identical with those in standard QED: with all particles on-shell, to satisfy energy--momentum conservation, this process must be absent. Then, for the effective QG scale entailed in the model, an order of magnitude, $\Ms/(\sigma\nd\ls)\sim 10^{17}$~GeV, c.f.~(\ref{eq:cdf}), is sufficient to make the D-foam interpretation of the light-speed variation effect viable, in which case the Cherenkov bound that would be imposed otherwise is irrelevant for the above reason.
\item In our previous foam model of~\cite{Li:2023wlo} the kinematics assumed in many phenomenological approaches to LV, including that of~\cite{He:2022jdl}, is inapplicable as energy is lost in the process. As we have already argued, it is such violations, exclusive to this approach, that raise the scale above which electron may decay:
\begin{equation}\label{eq:thrv3}
E>\frac{2m_{e}}{\sqrt{-\varDelta}},\quad\textrm{where}~\varDelta\coloneqq 2\Bigl(1-\frac{2\varsigma_{\I}{}^{\g}}{\Dg^{2}}\Bigr)\Delta v_{\g};
\end{equation}
or equivalently, for UHE electrons, $m_{e}\ll p\simeq E\geq E_{th}\equiv(4m_{e}^{2}\E_{\ast})^{1/3}$ from eq.~(\ref{eq:thrv1}). The new energy scale $\E_{\ast}$ we introduce here for $e\rightarrow e\g$ is given by
\begin{equation}\label{eq:nesd}
\E_{\ast}=\frac{E_{\textrm{LV}\g}}{1-2\varsigma_{\I}{}^{\g}E_{\textrm{LV}\g}/\Md},
\end{equation}
where $\Delta v_{\g}=-\Dg^{2}E_{th}/(2\Md)=-E_{th}/(2E_{\textrm{LV}\g})$ at threshold is used. So if the photon-velocity `anomaly' can finally be attributed to this type of foam, indicated by the last equality, the threshold limits as derived~\cite{He:2022jdl} for the case of ($\eta=0$, $\xi<0$), are actually imposed for $\varDelta$ or on the scale $\E_{\ast}$ in our case, though can be evaded if $\E_{\ast}\leq 0$ or easily satisfied in cases $\E_{\ast}>0$ via an assignment for the value of $\varsigma_{\I}{}^{\g}$ as above; they do \emph{not} constrain the \emph{actual} subluminal velocity $\Delta v_{\g}$ and the corresponding scale $E_{\textrm{LV}\g}$. The latter, related to $\Ms/(\Dg^{2}\gs)$, can only be limited in relation to time lags in the arrival of cosmic $\g$-rays as in eq.~(\ref{eq:cdf}).
\end{itemize}
Therefore, despite the fact that the nonobservation of electron decay via Cherenkov process imposed strict limits on subluminal LV effects for the photons, with a strength seven orders of magnitude~\cite{He:2022jdl} or eight tighter than the bound from time-of-flight analyses~(c.f.~(\ref{eq:cdf})), by this means one cannot exclude the similar effect associated with photon refraction induced by stringy space-time foam. The synchrotron radiation measurements also support D-foam models which induce \emph{no} LV in the charged lepton sector, provided from such observations, one could basically rule out subluminal LV effects in electron dispersion. These important features allow the hint of light-speed variation~\cite{Shao:2009bv,Li:2020uef} to survive the stringent Crab-Nebula constraints within such models.

We must emphasize once more, however, that any speculation about Lorentz violations in cosmic radiation needs support from more data. With many more UHE astrophysical $\g$-ray data being accrued by LHAASO and other experiments, it might be able to disentangle source from possible propagation effects due to fundamental~(string) physics. On the other hand, from the perspective of the D-particle models endorsed here, other cosmic probes like neutrinos may help to discriminate between  the above two approaches: the string-stretched linear-in-energy effects~(\ref{eq:sstd}),~(\ref{eq:sspri}), when applied to neutrinos, stay the same~\cite{Ellis:2008gg}, while for the models of~\cite{Li:2023wlo} they become entangled with the particle/antiparticle nature of neutrinos as evidenced from eq.~(\ref{eq:sfndr}). A more detailed recapitulation of their respective predictions can be found elsewhere~\cite{Li:2025stf}.

\section{\label{sec:concl}Discussion}

In this paper we have revisited electron Cherenkov effect \emph{in vacuo}, a nonstandard Lorentz-breaking effect for which we have no evidence till today, and thus, whose absence in particular has been used by some recent analyses~\cite{He:2022jdl,Li:2022ugz} to improve prior limits on certain types of Lorentz violations in QED. Clearly, the observational hint against such a process is indeed relevant for a pure-kinematics phenomenological formalism~(where just modified dispersion is postulated) as adopted therein, but in ways that it seems difficult to quantify, due to the dependence on the strength of interaction~(dynamics/matrix element), which is not known in this context. So, here, we have contemplated the implications of these findings from the perspective of the comments offered in the introduction section~\ref{sec:intro} concerning the plausibility~(and, practically, lack thereof) of conspiracies between modifications to kinematics and that of dynamics. We see that the latter is not particularly important as deriving the rates, and it is verified that in cases of our interest, above the threshold the rate is so tremendous that, it would allow a threshold analysis \emph{alone} with trans-Planckian sensitivity which relies on photon observations from distant nebulae such as the Crab.

We should mention that the effectiveness of an analysis of this type is somewhat limited by the fact that for these PeV gamma-rays LHAASO observed from the Crab, one can only reasonably guess a part of the source mechanisms for producing such high-energy emission. According to the most commonly adopted model~\cite{Atoyan:1996kqd}, the relevant gamma-rays are emitted as a result of the IC processes, and from this, one infers that for electrons of energies up to 2.3~PeV~\cite{LHAASO:2021cbz} the vacuum Cherenkov process is still ineffective. On the basis of the arguments just provided this then allows us to exclude certain corresponding regions of the parameter space by simply demanding~\cite{He:2022jdl,Li:2022ugz} $2.3~\textrm{PeV}\leq E_{e,th}$, for the thresholds~(c.f. eqs.~(\ref{eq:sct}),~(\ref{eq:hct})). Of course, apparently reliable constraints can be obtained by considering the highest energy measured directly for the electron/positron; in cosmic rays, energies of 40~TeV are reported recently by H.E.S.S.~\cite{HESS:2024etj}, although the resulting constraint is $10^{5}$ times weaker. Therefore, these kinematical~(threshold) constraints are fairly effective and in particular, the linear LV scale characterizing the \emph{subluminal} photon \emph{dispersion} is tightly bounded by the absence of the \emph{hard} Cherenkov effect by these particles.

However as it has become clear in section~\ref{sec:evs}, one cannot use this result to set constraints to certain models of~(stringy) QG, according to which effective subluminal refractive indices of a 4D nonstandard vacuum, as `seen' by the photons, may be induced, via specific types of reduced-Lorentz invariance. In particular, in section~\ref{sec:sstf}, we discuss such instances arising in brane-world scenario and we see that within D-foam models hereby provided albeit linearly modified photon propagation is found to characterize the situation electrons and, in general charged excitations, remain undisturbed by the presence of space-time D-brane defects. At first sight this is just the LV parameter configuration of $(\eta=0$, $\xi<0)$, allowed by a general setting~(\ref{eq:gdr}), pointing likely towards the possibility of hard Cherenkov effect. While in such noneffective field theories for QG foam, this is not the case. In subsection~\ref{sec:ace}, we described briefly the reasoning based on two specific foam situations, including the cases of stretched strings and isotropic, stochastic recoil fluctuations. In both cases, although fermions could travel faster than photons with comparable energies, at least effectively no such a radiation happens. Therefore in the contexts of such models one can develop a consistent description for the intriguing result of refs.~\cite{Shao:2009bv,Li:2020uef} where a linearly Planck-scale suppressed light-speed variation was suggested from analyzing travel times of GRB and AGN photons, and which may thus serve as a support to such string models~\cite{Li:2021gah}.

Before closing we would like to mention that there is another peculiar process permitted by general LV-deformed dispersions of particles, of the type~(\ref{eq:gdr}), namely the spontaneous decay of a single photon $\g\rightarrow ee^{+}$ that also occurs only above some threshold, $E_{\g,th}$, whose functional form, again, depends on hierarchies between parameters $\xi$ and $\eta$, as in the case of Cherenkov effect. It was well established~\cite{Mattingly:2005re,Jacobson:2005bg} that as $E_{\g}>E_{\g,th}$ its rate very rapidly becomes proportional to $E_{\g}^{2}/M$ such that, for a photon of order $E_{\g,th}\simeq 10$~TeV~(assuming the linear $\Mp$ suppression of the effect), the corresponding decay time is of order $\sim 10^{-10}$~s. In comparison with the observationally required timescale of $10^{11}$~s, i.e., the flight time for an observed Crab photon for which this reaction must not occur during its travel to Earth, constraints cast simply by requiring that the threshold itself is above the measured energy are also reliable. In contrast to the relativity violations that $e\rightarrow e\g$ can constrain, $\g$-decay will limit the \emph{superluminal} aspect of photon Lorentz violation and the \emph{subluminal} effects in electrons/positrons. Consider the simplest case(s), where LV is present only for one sector, the thresholds are translated into constraints on the violation scales: a) without any effect in the fermion dispersion relations~(i.e., setting $\eta=0$), from $E_{\g,th}$, of the form of eq.~(\ref{eq:hct}) by implementing $E_{e,th}\mapsto -E_{\g,th}$, we have~\cite{Li:2021tcw},
\begin{equation}\label{eq:pdc1}
E_{\textrm{LV}\g}\gtrsim 9.57\times 10^{32}~\textrm{eV}\,\Bigl(\frac{E_{\g}}{1~\textrm{PeV}}\Bigr)^{3},\quad\textrm{for}~s_{\g}=-1~(\textrm{c.f.}~\xi>0);
\end{equation}
on the contrary b) assuming Lorentz invariant photon $\xi=0$, but that the outgoing leptons are subject to~(subluminal) LV, from the corresponding threshold, $E_{\g,th}^{3}=-8m_{e}^{2}M/\eta$~(see, e.g.,~\cite{Jacobson:2002hd,He:2022jdl}) one obtains,
\begin{equation}\label{eq:pdc2}
E_{\textrm{LV}e}\gtrsim 4.79\times 10^{32}~\textrm{eV}\,\Bigl(\frac{E_{\g}}{1~\textrm{PeV}}\Bigr)^{3},\quad\textrm{for}~s_{e}=+1~(\textrm{c.f.}~\eta<0).
\end{equation}
Based on the first discovery of 12 Galactic PeVatrons by LHAASO~\cite{LHAASO:2021gok}, with observed energies up to the highest at 1.42~PeV, tight $\g$-decay limits were obtained from eqs.~(\ref{eq:pdc1}),~(\ref{eq:pdc2}) in the respective cases~\cite{Li:2021tcw,He:2022jdl}: a) $E_{\textrm{LV}\g}>2.7\times 10^{24}$~GeV; b) $E_{\textrm{LV}e}>1.4\times 10^{24}$~GeV. In the former case the collaboration implied a similar constraint~\cite{LHAASO:2021opi}.

To make our paper most useful for the reader who looks for the severest bound ever to this phenomenon, and hence on these two types of modifications in particle dispersions, we note that these scales can be further constrained with LHAASO's latest report on an UHE $\g$-ray \emph{bubble}~\cite{LHAASO:2023uhj} in the direction of the active star-forming region Cygnus-X, the region from which the previous $\sim 1.4$~PeV photon event was found. Gamma-rays obtained towards the Cygnus bubble are observed up to an unprecedented high energy of 2.48~PeV.\footnote{We note that the detected UHE photons spread over the entire bubble region make unlikely the leptonic IC origin of the radiation and indicate the involvement of hadronic process~\cite{LHAASO:2023uhj}, so that the argument leading to electron Cherenkov bound as applied to the Crab is not valid in this case.} So, we find the constraints:
\begin{align}\label{eq:pdc3}
E_{\textrm{LV}\g}\geq 1.5\times 10^{25}~\textrm{GeV}&\quad\Leftrightarrow\quad\xi\leq 8.4\times10^{-7};\\
\label{eq:pdc4}
E_{\textrm{LV}e}\geq 7.3\times 10^{24}~\textrm{GeV}&\quad\Leftrightarrow\quad\eta\geq -1.7\times10^{-6}.
\end{align}
The estimates we make tighten the strongest one~(among the existing values, say of refs.~\cite{LHAASO:2021opi,Li:2021tcw,He:2022jdl}) by a factor of $\sim (12/7)^{3}\approx 5$. The inclusion of more general cases having $\xi\neq 0$ and $\eta\neq 0$, is straightforward, and, parallels the analyses carried out in~\cite{He:2022jdl} using 1.4~PeV as the greatest photon energy, prior to the discovery of the $\sim 2.5$~PeV event. While an analysis in this line does not present any technical difficulty, it requires a separate investigation which takes us out of the purpose of the present paper.

Finally, one should observe that the trans-Planckian lower limits for these types of LV, or equivalently, the small values of the corresponding dimensionless parameters $\xi$ and $\eta$ are consistent with the \emph{subluminal} feature of the D-foam-induced vacuum refractive indices for photons, capable of accommodating the hint of light-speed variation exposed in the time-of-flight tests~\cite{Shao:2009bv,Li:2020uef}. Indeed in the context of such brane models, as advocated here, photons are stable~(i.e., do \emph{not} decay), so that~\cite{Li:2021gah} the strong $\g$-decay constraints thereof~\cite{LHAASO:2021opi,Li:2021tcw,He:2022jdl}, as well as the most stringent result just derived, are inapplicable. For reconciliation of such models with LHAASO data, especially in this respect, we refer the reader to our paper~\cite{Li:2021tcw} for details. Moreover, it should be noted that the peculiarity of the transparency of D-foam to electrons or charged probes, for specifically stringy properties outlined above,\footnote{It is for the same reason that such models also survive the severe constraints~(see, e.g.,~\cite{He:2024ljr}) that would be obtained from models with LV modifications in protons and/or pions and thus modifications in hadronic GZK cutoff, which is unaffected due to vanishing or suppressed interactions between particles involved~(i.e., protons, or pions whose constituents are charged) and D-foam.} provides additional reasons for escaping the otherwise very stringent bounds for putative LV effects on electrons/positrons, including~(\ref{eq:pdc4}). For this latter point, see also~\cite{Li:2022ugz}. This is similar to the way, discussed in that paper, out of electron~(soft) Cherenkov limits on this parameter imposed from the positive side.

\acknowledgments
C.L. is indebted to an anonymous reviewer of his Ph.D. Thesis for useful remarks, regarding foam effect on charged leptons, that have partially inspired the work presented in this paper. This work is supported by the National Natural Science Foundation of China under Grants No.~12335006 and No.~12075003. The work of C.L. is partly supported by a \emph{Boya} Fellowship provided by Peking University, and by the China Postdoctoral Science Foundation~(CPSF) under Grant No.~2024M750046. C.L. gratefully acknowledges the financial support received from Postdoctoral Fellowship Program~(Grade B) of CPSF under Grant No.~GZB20230032. The work of C.L. also falls within the scope of the COST Action CA23130 -- Bridging high and low energies in search of quantum gravity~(BridgeQG).

\appendix

\section{Implication of brane recoil on space-time}
\label{app}

Formally space-time foam situations in the cosmological formalism we use~\cite{Ellis:2004ay} are described by nonconformal deformations of a fixed-point $\sigma$-model $S_{0}$ with vertex LCFT operator $V_{g^{J}}$'s on the worldsheet $\Sigma$~($X^{\mu}(\sigma,\tau)$ are target-space coordinates/matter fields):
\begin{equation}\label{eq:app1}
S_{m}=S_{0}+g^{J}\int_{\Sigma}V_{g^{J}}(X),
\end{equation}
which in a first-quantized formalism, governs the dynamics of matter excitations of a string Universe in a near equilibrium situation, where fluctuations of foam vacua are assumed~(for consistency of such a perturbative treatment) to be dilute. Such is the case of back-reaction of D-particle recoil onto space-time, a subject that shall be reviewed in this appendix based on~\cite{Kogan:1996zv,Ellis:1996lrt} where the reader is referred for further details.

Consider a D-particle~(0-brane), initially \emph{at rest} at $Y^{i}(t=0)\equiv Y_{i}$ of spatial coordinates of a $(D+1)$-dimensional space-time~(which could be a 3-brane-world), experiencing at $t=0$ an \emph{impulse} due to scattering off a matter string. For a heavy defect~\cite{Ellis:1999uh} its trajectory, $Y^{i}(t)$, is specified by~(\ref{eq:app1}), where the vertex operator, corresponding to the present case with the relevant background $g^{J}=u_{i}$, is,
\begin{equation}\label{eq:app2}
V_{imp}=\int_{\partial\Sigma}\varepsilon(\varepsilon Y_{i}+u_{i}X^{0})\Theta_{\varepsilon}(X^{0})\partial_{\perp}X^{i},
\end{equation}
where $\partial_{\perp}$ is the normal derivative on the worldsheet boundary $\partial\Sigma$, and, $X^{i}$, $X^{0}$~(whose zero mode is the target time, $t$), are $\sigma$-model fields, satisfying Dirichlet and Neumann boundary conditions respectively. The recoil/impulse of the D-particle, turned on `abruptly' at $t=0$, is realized by promoting the Heaviside $\Theta$-function to a \emph{regularized} operator:
\begin{equation}\label{eq:app3}
\Theta_{\varepsilon\rightarrow 0^{+}}(t)=-\frac{\i}{2\pi}\int_{-\infty}^{\infty}\frac{\sd q}{q-\i\varepsilon}\e^{\i qt}.
\end{equation}
Eq.~(\ref{eq:app2}) contains actually a \emph{pair} of deformations, which are relevant in a worldsheet renormalization group~(RG) sense, having anomalous dimensions $-\varepsilon^{2}/2$, where $\varepsilon$ is a regulator. They form a modified~(logarithmic) conformal algebra~(superconformal algebra, in the case of superstrings), which \emph{closes} if and only if, one links~\cite{Mavromatos:1998nz} $\varepsilon^{-2}$ with the `running' RG scale, $\ln A$, given in terms of the worldsheet area $A$:
\begin{equation}\label{eq:app4}
\varepsilon^{-2}\sim\ln A.
\end{equation}

Upon the identification~(\ref{eq:app4}) the~(renormalized/rescaled) couplings $Y_{i}=\frac{\varepsilon Y_{i}}{\varepsilon}$ and $u_{i}=\frac{\varepsilon u_{i}}{\varepsilon}$ are \emph{marginal}, that is, independent of the scale, $\varepsilon$. It is these marginal couplings that are connected to target-space quantities of physical relevance, such as space-time back reaction of recoil, resulting in a sort of Finsler-type metric distortion~(\ref{eq:std}), which we now proceed to discuss, and thus of the associated refraction/delay in the propagation of neutral matter or radiation, c.f.~(\ref{eq:dfptd}) in the text. Note that in the long-time limit, i.e., $\varepsilon\rightarrow 0^{+}$, the dominant contributions come from $u_{i}$ recoil deformation, in~(\ref{eq:app2}).~(For formal reasons of convergence of worldsheet path integral the space-time shall be assumed Euclidean, in which case LCFT of recoil~\cite{Kogan:1996zv,Gravanis:2002ii} is rigorously valid, however the final dressed space-time~(\ref{eq:std}) can eventually acquire Minkowski signature~\cite{Gravanis:2002gy}.)

The introduction of such a deformation violates conventional conformal symmetry, as a modified one~\cite{Mavromatos:1998nz} holds. The theory is actually \emph{supercritical}, due to a central charge \emph{surplus}, $Q^{2}\sim \varepsilon^{4}/\gs^{2}>0$~(compared to the critical-string value fixed by equilibrium situations) and, requires gravitational dressing via a \emph{Liouville field}, $\varphi$~\cite{Ellis:1992eh,Ellis:1999uh}: $V_{imp}\rightarrow V_{imp}^{L}$~(for consistency of the worldsheet theory). This results in the recoil $u_{i}$ term in eq.~(\ref{eq:app2}) being replaced by a bulk operator~($\alpha=1,2$ is a worldsheet index):
\begin{align}\label{eq:app5}
V_{imp}^{L}\rightarrow V_{bulk}\sim&\int_{\Sigma}\e^{\alpha_{\pm}\varphi}\partial_{\alpha}(\varepsilon u_{i}t\Theta_{\varepsilon}(t)\partial^{\alpha}X^{i})\nonumber\\
\supset& -\varepsilon^{2}u_{i}\int_{\Sigma}\e^{\varepsilon(\varphi-t)}t\theta(t)\partial_{\alpha}\varphi\partial^{\alpha}X^{i},\quad\textrm{for}~t>0,
\end{align}
where $\alpha_{\pm}\sim\pm\varepsilon$ is the gravitational conformal dimension. The supercriticality of the string thereby, implies a time-like signature for the $\varphi$ field, whose zero mode is identified with the target Minkowski time $t$ for $t\rightarrow\infty$ after the impulse inducing \emph{locally} off-diagonal terms in the final Liouville-augmented space-time~(prior to averaging over D-particle populations in the foam)~\cite{Ellis:1999uh,Kogan:1996zv,Ellis:1996lrt,Ellis:1997cs}:
\begin{equation}\label{eq:app6}
G_{\mu\nu}=\eta_{\mu\nu}+h_{\mu\nu},\quad h_{0i}\simeq u_{i\Vert}=\gs\frac{\Delta p_{i}}{\Ms},
\end{equation}
that is eq.~(\ref{eq:std}), where the suffix 0 denotes the temporal~(Liouville) component, and $u_{i}$, the D-particle recoil velocity, is related to the momentum transfer $\Delta\pv$ as a result of momentum conservation in each scattering event.

The reader should recall that the velocity dependence of the metric~(\ref{eq:app6}) and hence the dependence of the latter on the momentum of the particle make such metrics quite different from usual metric perturbations, that depend only on coordinates, in particular concerning the fact, that~(\ref{eq:app6}) can be diagonalized with a T-duality transformation of coordinates~\cite{Mavromatos:2010pk}. As already mentioned, the recoil of a D-particle resembles the case of strings propagating in a target-space electric $B$ field with $B_{0i}\sim u_{i}$. This leads to mixed-type boundary conditions on the boundary of worldsheet surfaces with topology of a disc:
\begin{equation}\label{eq:app7}
G_{\mu\nu}\partial_{\perp}X^{\nu}+B_{\mu\nu}\partial_{\tau}X^{\nu}\vert_{\partial\Sigma}=0.
\end{equation}
Consider a frame where the particle has motion only in the spatial $X^{1}$ direction the induced target-space metric, felt by the~(electrically neutral) string matter, is obtained by evaluating the relevant worldsheet propagator:
\begin{align}\label{eq:app8}
&G_{\mu\nu}=\bigl(1-u^{i}u_{i}\bigr)\eta_{\mu\nu},\quad\ \ \mu,\nu=0,1,\nonumber\\
&G_{\mu\nu}=\eta_{\mu\nu},\quad\mu,\nu=\textrm{all other values}.
\end{align}
Even in diagonal form~(\ref{eq:app8}) the Finslerian nature~\cite{Bao:2000rfg} of the metrics differentiates them significantly from metric perturbations used in relativity, due to the presence of a $u$-dependent piece. Also, the form of this metric implies the existence of a \emph{critical} recoil velocity, i.e., the low-energy speed of light \emph{in vacuo}, related implicitly to Lorentz symmetry of the underlying string theory~\cite{Ellis:2010he}.

It is important to notice that deformations induced by different D-particles `sum' up all together: a \emph{uniformly averaged} effect of the foam then occurs from a macroscopic viewpoint. In a randomly fluctuating D-particle recoil model such as that of~\cite{Li:2023wlo}, the metric of~(\ref{eq:app6}) is further taken to be stochastic, Gaussian~(owing to a lack of any particular prior knowledge), i.e., the corresponding averaging of a quantity ${\cal Q}(\zeta_{\Vert})$ over this foam entails integration over the statistical variable $\zeta_{\Vert}$ through integrals of the form
\begin{equation}\label{eq:app9}
\frac{1}{\D\sqrt{2\pi}}\int_{-\infty}^{\infty}\sd\zeta_{\Vert}{\cal Q}\exp\biggl(-\frac{\zeta_{\Vert}^{2}}{2\D^{2}}\biggr),\quad\textrm{with}~\sdla\zeta_{\Vert}^{2}\sdra=\D^{2},
\end{equation}
in which case there is a \emph{statistical Lorentz symmetry} despite stochastic $u_{i}$ fluctuations which feed the effects of photon refractive indices of~(\ref{eq:sfpri}). Moreover, there are losses in the total energy in particle interactions, due to the presence of foam such that the anomalous~(hard) Cherenkov effect \emph{in vacuo} by the UHE electrons~($p\gtrsim 10^{15}$~eV) which go slightly faster than photons in such a quantum-gravitational environment, is suppressed. It is in this way, that the otherwise stringent constraints as developed in~\cite{He:2022jdl,Li:2022ugz} are avoided.

Envisaging stretched-string mechanism of delay and thus of the associated subluminous photon refraction by the nontrivial vacuum constitutes a truly microphysical model of foam, where the capture/splitting of the string creates quantum oscillating, nonlocal intermediate string states, associated with the stringy uncertainties; this has been briefly reviewed in the text and discussed in detail in~\cite{Ellis:2008gg}. In such scenarios, there would also be compatibility with the current observational hint of a Lorentz-violating speed variation in cosmic photons~\cite{Shao:2009bv}. In that case, as we pointed out in our work, given the disentanglement of the induced casual delays from the corresponding dispersion relations of photons the usual Lorentz kinematics for the electrons~(positrons) is easily guaranteed.

\end{document}